\documentclass[aip,rsi,reprint,superscriptaddress]{revtex4-1}
\usepackage{graphicx}

\begin{document}

%\preprint{}

\title{Single Spin Optically Detected Magnetic Resonance with E-Band Microwave Resonators.}


%\author{}
%\thanks{}
%\affiliation{}

\author{Nabeel Aslam}
\email[]{n.aslam@physik.uni-stuttgart.de}
%\thanks{}
\author{Matthias Pfender}
\author{Rainer St\"ohr}
\author{Philipp Neumann}
\affiliation{3. Physikalisches Institut, University of Stuttgart, Pfaffenwaldring 57, 70569 Stuttgart, Germany}
\author{Marc Scheffler}
\affiliation{1. Physikalisches Institut, University of Stuttgart, Pfaffenwaldring 57, 70569 Stuttgart, Germany}
\author{Hitoshi Sumiya}
\affiliation{Sumitomo Electric Industries Ltd., Itami, 664-001, Japan}
\author{Hiroshi Abe}
\author{Shinobu Onoda}
\author{Takeshi Ohshima}
\affiliation{Japan Atomic Energy Agency, Takasaki, 370-1292, Japan}
\author{Junichi Isoya}
\affiliation{Research Center for Knowledge Communities, University of Tsukuba, Tsukuba, 305-8550 Japan}
\author{J\"org Wrachtrup}
\affiliation{3. Physikalisches Institut, University of Stuttgart, Pfaffenwaldring 57, 70569 Stuttgart, Germany}


\date{\today}

\begin{abstract}
Magnetic resonance with ensembles of electron spins is nowadays performed in frequency ranges up to $240\,$GHz and in corresponding magnetic fields of up to $10\,$T.
However, experiments with single electron and nuclear spins so far only reach into frequency ranges of several $10\,$GHz, where existing coplanar waveguide structures for microwave (MW) delivery are compatible with single spin readout techniques (e.g. electrical or optical readout).
Here, we explore the frequency range up to $90\,$GHz, respectively magnetic fields of up to $\approx 3\,$T for single spin magnetic resonance in conjunction with optical spin readout.
To this end, we develop MW resonators with optical single spin access.
In our case, rectangular E-band waveguides guarantee low-loss supply of microwaves to the resonators.
Three dimensional cavities, as well as coplanar waveguide resonators enhance MW fields by spatial and spectral confinement with a MW efficiency of $1.36\,\mathrm{mT/\sqrt{W}}$.
We utilize single NV centers as hosts for optically accessible spins, and show, that their properties regarding optical spin readout known from smaller fields ($<0.65\,$T) are retained up to fields of $3\,$T.
In addition, we demonstrate coherent control of single nuclear spins under these conditions.
Furthermore, our results extend the applicable magnetic field range of a single spin magnetic field sensor.
Regarding spin based quantum registers, high fields lead to a purer product basis of electron and nuclear spins, which promises improved spin lifetimes.
For example, during continuous single-shot readout the $^{14}$N nuclear spin shows second-long longitudinal relaxation times.
\end{abstract}

%\keywords{}

\maketitle

\section{Introduction}
Single electron and nuclear spin magnetic resonance is routinely detected on a variety of systems nowadays, ranging from single molecules\cite{wrachtrup_optical_1993,wrachtrup_optically_1993}, solid state defects\cite{jelezko_observation_2004,weber_defects_2011,pla_high-fidelity_2013} to rare earth ions\cite{siyushev_coherent_2014}.
In all these cases, spin control is performed via oscillating magnetic fields, but readout is performed either electrically or optically.
For small magnetic fields and hence moderate microwave (MW) frequencies up to several $10\,$GHz, this is compatible with wires, antennas and coplanar waveguide (CPW) structures manufactured close to the spin \cite{toyli_chip-scale_2010,pla_high-fidelity_2013}.
At higher MW frequencies the transmission-loss of coaxial cables increases drastically.
The main reason is the decreasing skin depth $\propto$ 1/$\sqrt{f}$, which leads to higher effective resistivity in the conductor line.
Increasing the diameter of the coaxial cables to compensate this effect is only possible up to a certain point, since it decreases the moding-frequency.
Additionally, at higher frequencies (e.g. in the E-band from $60\,$GHz to $90\,$GHz) semiconductor-based MW sources and amplifiers currently achieve an output power of less than $28\,$dBm.
Taking this into account, novel strategies for combining single spin magnetic resonance at high frequencies and specific readout techniques need to be developed that ensure a high microwave-power-to-magnetic-field conversion efficiency $C_{\text{mag}}= B_1/\sqrt{P_{\mathrm{MW}}}$ (respectively microwave-power-to-Rabi-frequency conversion efficiency $C_{\text{Rabi}}= \Omega/\sqrt{P_{\mathrm{MW}}}$).
\\
\indent
Working in the E band extends the accessible range of single spins as magnetic field sensors up to $3\,$T.
Furthermore, experiments at such high bias magnetic field promise additional striking improvements.
For spin based quantum information candidate systems like phosphorous in silicon (Si:P)\cite{pla_high-fidelity_2013,morello_single-shot_2010}, rare-earth ions\cite{siyushev_coherent_2014,kolesov_mapping_2013,kolesov_optical_2012,yin_optical_2013} or nitrogen-vacancy (NV) centers in diamond\cite{waldherr_quantum_2014,dolde_high-fidelity_2014,togan_quantum_2010,bernien_heralded_2013-2} higher fields reduce dressing among electron and nuclear spin states.
Hence, energy eigenstates converge to pure product spin states, which promises improved coherence properties and nuclear spin access\cite{mccamey_electronic_2010,waldherr_quantum_2014}.
The single spin sensors (e.g. NV centers in diamond\cite{balasubramanian_nanoscale_2008,maze_nanoscale_2008}) can be applied for nanoscale detection of external electron and nuclear spins\cite{haberle_nanoscale_2015-1,mamin_nanoscale_2013,devience_nanoscale_2015,staudacher_nuclear_2013,grotz_sensing_2011,grinolds_nanoscale_2013}.
High fields are particularly important for nuclear spin detection because under these conditions chemical shifts of nuclear spin resonances allow chemically specific sensing in homonuclear environments.
Finally, the thermal spin polarization increases with higher available magnetic fields.
\\
\indent
Here, we develop and demonstrate a single spin magnetic resonance setup with optical spin readout (optically detected magnetic resonance -- ODMR) for magnetic fields up to $3\,$T.
This method is in principle also applicable to higher magnetic fields.
As a single spin test system we use nitrogen-vacancy (NV) color centers in diamond\cite{gruber_scanning_1997}.
\\
\indent
The electron spin of the NV center is a triplet ($S=1$) with a zero field splitting of $D=2.87\,$GHz.
An external magnetic field aligned along the NV axis lifts the degeneracy of the $m_S=\pm1$ energy levels.
The latter spin states split with $2\cdot 28.03\,\mathrm{MHz / mT}$.
Single NV centers are optically addressed via confocal microscopy, and MW fields in the E-band are delivered via respective rectangular waveguides (WR-12).
At the position of the NV center, we fabricate waveguide and CPW resonators for spatial and spectral field concentration.
Single nuclear spins are controlled via radio-frequency (RF) fields delivered by wires and read out by electron nuclear double resonance (ENDOR) via the NV center electron spin.

\section{single spin ODMR setup up to 3T}
\begin{figure*}[t]
	\begin{center}
	\includegraphics[width=\linewidth]{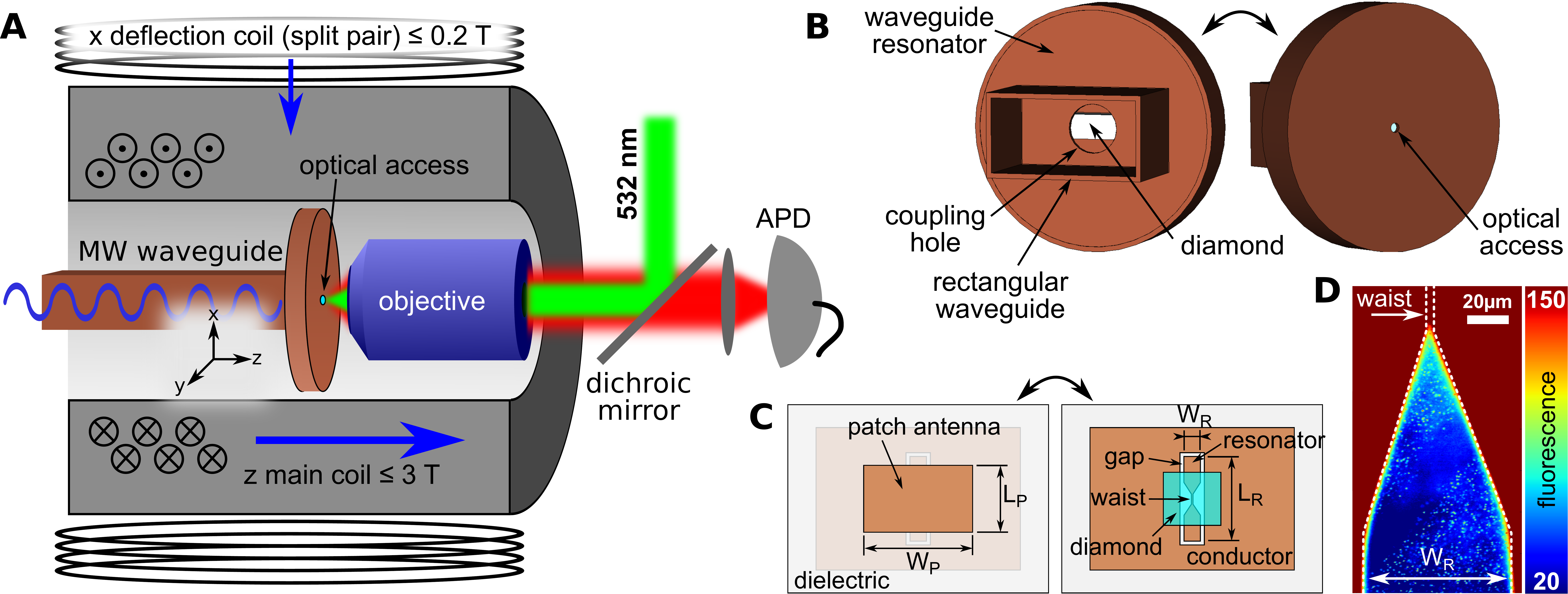}
	\caption
	{
		\textbf{Experimental setup.}
		\textbf{A,} bore of a $3\,$T, $0.2\,$T, $0.2\,$T superconducting vector-magnet with a bore diameter of $10\,$cm.
		The direction of the magnet's main coil coincides with the optical axis.
		Optical excitation is performed by $532\,$nm laser light being focused through a $525\,\mu$m diameter hole onto the diamond located inside the microwave cavity.
		The resulting red-shifted NV fluorescence is separated from the excitation beam by a dichroic mirror and imaged onto an APD chip.
		For visibility reasons, the microwave and optical components are enlarged and the second deflection coil is omitted.
		\textbf{B,} schematic of the TM$_{110}$ MW cavity resonator.
		On the left, the rectangular waveguide providing the MW excitation is outlined.
		Inductive coupling is achieved by a small hole in the middle of the resonator plate facing the waveguide.
		The right side shows a small hole providing optical access, behind which the diamond is located.
		\textbf{C,} schematic of the waveguide to CPW resonator transition element.
		It consists of a dielectric layer, with the patch antenna facing the rectangular waveguide on one side (left image) and the CPW resonator structure on the other (right image). The diamond is mounted on top of the resonator.
		\textbf{D,} confocal scan of a diamond sample mounted on top of the CPW resonator.
		Since the dielectric layer is fluorescent, the antenna as well as the tapering is visible.
		In areas with low fluorescence background, fluorescent spots corresponding to single NV centers are visible.
		\label{fig:1}
	}
	\end{center}
\end{figure*}
At the heart of our single spin ODMR setup there is the confocal microscope which focuses $532\,$nm excitation laser light through an oil immersion objective onto an NV center in diamond (see figure~\ref{fig:1}).
The red-shifted fluorescence response is spectrally and spatially filtered and finally imaged onto a single photon counting module.
We utilize a home-built microscope which fits in the room temperature bore (diameter $10\,$cm) of a superconducting magnet with up to $3\,$T main field and up to $0.2\,$T perpendicular deflection coils (Scientific Magnetics).
We have placed sample positioners and the objective into the bore, all other optical, MW and RF elements are placed outside the magnet.
The main magnetic field axis coincides with the optical axis, optimizing the setup for (111)-oriented diamond substrates with NV centers pointing along the surface normal.
\\
\indent
For this work a (111) diamond plate was cut from a low-strain type-IIa high-pressure-high-temperature (HPHT) crystal with an average substitutional nitrogen concentration of $11\,$ppb and natural abundance of $^{13}$C.
The plate was irradiated with $2\,$MeV electrons to a total dose of $2\cdot10^{10}\,\mathrm{e/cm^2}$ and was subsequently annealed at $1000\,^{\circ}$C for 2 hours in vacuum.
Then, a thin film of $1\,\text{mm}\times 1\,\text{mm}\times 90\,\mu$m was obtained by laser-cutting and polishing from the plate.
\\
\indent
For the generation of MW we combine an Anritsu signal generator MG3690C with an OML S12MS frequency multiplier (6x) system with a measured output power of $0\,$dBm.
In the frequency range from $71\,$GHz to $76\,$GHz we employ a MW power amplifier (Sage Millimeter Inc.) with an output power of $20\,$dBm (P-1dB).
Microwaves in the E band are guided to the NV center via a rectangular waveguide (WR-12).
At the end of the waveguide we concentrate the microwave radiation spectrally and spatially to a region of diamond containing NV centers.
\\
\indent
We have implemented two different approaches for microwave delivery into the diamond: A three dimensional waveguide cavity and a transition from rectangular waveguide to CPW resonator (see figure~\ref{fig:1}).
As a first approach, we use a circular TM$_{110}$ cavity attached to the rectangular waveguide as depicted in figure~\ref{fig:1}B.
The diamond is placed inside this resonator on the face opposite to the coupling hole.
The latter opposite metal face has a hole for optical access.
Inside the cavity a MW field $B_1$ builds up, which is perpendicular to the bias field and leads to a Rabi frequency of the NV electron spin of $\Omega = - g_{\text{NV}}\cdot \mu_B \cdot B_1/\sqrt{2}$, with the NV electron g-factor $g_{\text{NV}}$ and the Bohr magneton $\mu_B$.
\\
\indent
The tailored transition from waveguide to CPW resonator (see figure \ref{fig:1}C) was inspired by low-loss microstrip line to waveguide transitions in automotive radar systems\cite{iizuka_millimeter-wave_2002}.
In contrast, our goal is to concentrate the MW field at the end of the waveguide.
To this end we design an assembly containing a tapered $\lambda/2$ dipole antenna ($f\approx 75\,$GHz).
Close to the waist position, MW radiation is spectrally and spatially confined and electron spins can be coherently manipulated.
Optical access is possible above the CPW resonator or inside the gaps.
\\
\indent
In addition to high MW efficiencies, both mentioned approaches should provide a linewidth of $\sim 100\,$MHz.
This allows addressing proximal nuclear spins with hyperfine couplings up to $\sim 100\,$MHz\cite{mizuochi_coherence_2009}.
In addition, fast gate operation times are not limited by a narrow resonator linewidth.
\\
\indent
RF is supplied by a wire across the CPW resonator perpendicular to the dipole axis to reduce losses (see figure~\ref{fig:4}B).

\section{Microwave resonators --- design and simulations}

\subsection{TM$_{110}$ waveguide cavity}
\begin{figure}[t]
\includegraphics[width=\columnwidth]{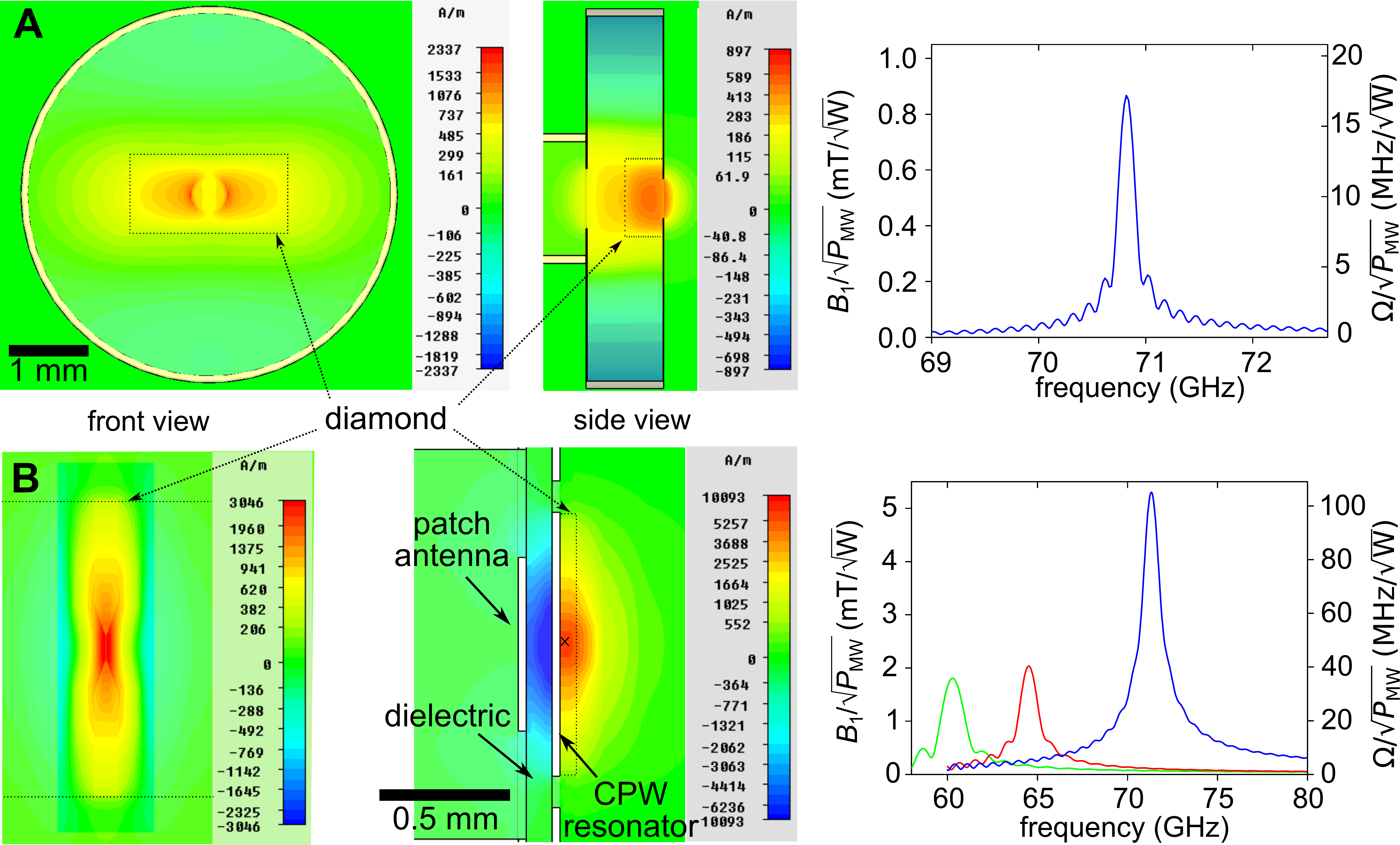}
\caption{\label{fig:2}
	\textbf{MW resonators in the E band - simulation results.}
	\textbf{A,} ``CST microwave studio'' simulation results for the TM$_{110}$ cavity.
	The left side shows the resonant $H_{1,x}$ field profile in the xy plane $20\,\mu$m inside the cavity.
	In the center, the corresponding field profile in yz direction is shown.
	On the right, the $C_{\text{mag}}$ (and $C_{\text{Rabi}}$) spectrum at a position $20\,\mu$m inside the cavity and in the center of the xy plane is plotted.
	The peak in the spectrum corresponds to the TM$_{110}$ mode.
	\textbf{B,} Results of simulation of the waveguide to CPW resonator transition.
	The lateral $H_{1,x}$ field profile is shown on the left for the xy, and in the middle for the yz plane.
	The green and red resonances on the right correspond to resonator \#1, the blue resonance to resonator \#2.
	For the blue and the green resonance, the diamond is centered above the antenna, while it is shifted away from the waist for the red curve.
	The $H_{1,x}$ field was evaluated $20\,\mu$m above the CPW, as indicated by the cross in the field profile.
 }
\end{figure}

The first approach to increase the MW field at the NV center location is a waveguide cavity with a suitable mode.
For the selection of the cavity mode, the orientation of the external magnetic field and the optical axis play an important role.
The latter rules out TE$_{011}$ mode resonators often used in EPR because of the inappropriate $B_1$ field profile,which is concentrated in the center of the cylindrical MW cavity where, consequently, the sample has to be mounted.
Our microscope objective has a working distance of $\approx 0.3\,$mm.
Contrary to this, a typical TE$_{011}$ cavity working at $75\,$GHz has a radius of $2.4\,$mm and length of $4.8\,$mm making an optical access unfeasible\cite{pozar2004microwave}.
\\
\indent
Our cavity mode of choice is the TM$_{110}$ mode.
In the proper configuration, its concentrated $B_{1}$ field has an orientation perpendicular to one of the crystallographic NV orientations in a diamond with (111) surface.
Inductive coupling from the WR-12 waveguide to the cavity is accomplished by a centered hole on the circular surface of the cavity facing the waveguide.
The orientation of the incoming $B_{1}$ field from the waveguide matches the orientation of the $B_{1}$ field of one out of two nominally degenerate TM$_{110}$ modes in the cavity with perpendicular polarizations, (see figure \ref{fig:2}) resulting in efficient coupling\cite{pozar2004microwave}.
$B_{1}$ is maximum in the center of the cavity and in transverse direction to the bias magnetic field (see figure \ref{fig:2}A).
The field has no node in longitudinal direction.
Hence, a centered hole in the second circular surface provides an optical access to the area with a highly concentrated $B_{1}$ field.
In addition, the higher refractive index of diamond compared to air leads to a further $B_1$ field enhancement in the diamond.
The cavity frequency is defined by the radius $a$ of the cavity.
Decreasing the cavity length $d$ has no effect on the resonance frequency, but increases the ratio of quality factor $Q$ and cavity volume $V$ \cite{pozar2004microwave}.
This ratio is approximately doubled for $2a/d = 5$ compared to the case of $2a/d = 1$, resulting in a higher $B_{1}$ field strength.
\\
\indent
By varying the diameter of the coupling hole, one can tune the regime between an undercoupled, a critically coupled and an overcoupled cavity.
In our case, a critically coupled cavity results in maximum $B_{1}$ at the NV center location.
Simulations were performed in order to estimate the optimum coupling hole diameter, predicting a cavity quality factor of $Q_{\text{c,sim}}=643$.
$20\,\mu$m below the optical access a maximum MW efficiency of $C_{\text{mag}}=0.87\,\mathrm{mT/\sqrt{W}}$ ($C_{\text{Rabi}}=17.6\,\mathrm{MHz/\sqrt{W}}$) is expected.
\\
\indent
The TM$_{110}$ cavity is assembled out of three separate parts.
The main component is a copper plate with a round bore and a thickness of $1\,$mm.
The second part is a copper film with a hole for inductive coupling to the cavity.
The cavity is closed on the other side by a glass slide with a thin copper layer comprising a hole for optical access ($525\,\mu$m diameter).
All parameters of the cavity are listed in table~\ref{tab:resonators}.
\\
\indent
One of the loss channels is conductive loss.
There is a non-zero current density across the contact area between the circular surfaces and the cylindrical part, leading to conductive losses at the edges of the cavity due to non-ideal contacts.
Furthermore, surface roughness is a crucial point which is improved by polishing the copper plate.
\\
\indent
Radiative losses also have to be considered.
Because the optical access in our case is almost as large as the hole for inductive coupling, the accompanied radiative loss is not negligible.
Moreover, a large optical access decreases the $B_1$ field in the region of interest (see the xy plot of the $H_{1,x}$ field in figure~\ref{fig:2}A).
%\out{Because of the optical access right at the position of the maximum $B_{1}$ field strength, we expect the radiative losses to be significant, as can be seen in the xy plot of the $H_{1}$ field.}
In fact, the MW efficiency increases to $C_{\text{mag}}=1.68\,\mathrm{mT/\sqrt{W}}$ ($C_{\text{Rabi}}=33.3\,\mathrm{MHz/\sqrt{W}}$) at the aforementioned position by removing the optical access in the simulation.
%\comment{this is not really due to radiative losses but due to a missing current density close to the NVs. Of course this is related to radiative losses.}
\\
\indent
In principle, other cavity modes can be used as well.
In the case of the TM$_{010}$, the $B_{1}$ field is maximum at the edges of the cavity where both the optical microscopy and the MW coupling from the waveguide would have to be placed\cite{pozar2004microwave}.
Another possibility is to shape the diamond itself, according to the corresponding cavity parameters, and cover it with metal.
In this case, the conductive losses at the edges and the mode volume would be decreased.
\\
\indent
The metal layer where the diamond is attached to has a thickness of $8\,\mu$m, which is above the skin depth of the applied MW radiation ($\approx 240\,\mathrm{nm}\,@\,75\,$GHz), but in the range of the skin depth ($\approx 6.5\,\mu$m) for nuclear spin resonance frequencies of $\approx 100\,$MHz.
Hence, a wire outside of the resonator can be used to deliver RF for coherent nuclear spin control for a proper metal layer thickness.
\begin{table}
\caption{Summary of geometric parameters of waveguide cavity and CPW resonators.\label{tab:resonators}}
\begin{ruledtabular}
		\begin{tabular}{l l l}
		\multicolumn{3}{c}{TM$_{110}$ cavity} \\
		\hline
    \multicolumn{2}{l}{diameter} & $4.8\,$mm \\
		\multicolumn{2}{l}{length} & $1.0\,$mm \\
		\multicolumn{2}{l}{diameter of optical access} & $525\,\mu$m \\ 
		\multicolumn{2}{l}{diameter of MW coupling hole} & $800\,\mu$m \\ 
		\multicolumn{2}{l}{thickness of metal face with optical access} & $8\,\mu$m \\ 
		\multicolumn{2}{l}{thickness of metal face with MW coupling hole} & $20\,\mu$m \\ 
		\multicolumn{2}{l}{material} & copper \\
		\hline
		Waveguide to CPW resonator &\#1&\#2\\
    \hline
		patch antenna length $\mathrm{L_P}$ & $729\,\mu$m & $667\,\mu$m \\ 
    patch antenna width $\mathrm{W_P}$ & $1531\,\mu$m & $1476\,\mu$m  \\ 
    resonator length $\mathrm{L_R}$ & $1085\,\mu$m & $1014\,\mu$m \\ 
		resonator width $\mathrm{W_R}$& $165\,\mu$m & $85\,\mu$m \\ 
    relative dielectric constant $\epsilon$ & \multicolumn{2}{c}{2.9} \\ 
    dielectric thickness & \multicolumn{2}{c}{$100\,\mu$m} \\ 
    conductor thickness & $10\,\mu$m & $30\,\mu$m \\
    gap CPW & $42\,\mu$m & $121\,\mu$m \\
    width of waist & $85\,\mu$m & $3\,\mu$m \\
    length of waist & $88\,\mu$m & $80\,\mu$m \\
		conductor material & \multicolumn{2}{c}{copper}\\
		dielectric material & \multicolumn{2}{c}{Rogers Ultralam 3850}
    \end{tabular}
\end{ruledtabular}
\end{table}

\subsection{Waveguide to coplanar resonator transition}
Further enhancement of the RF magnetic field amplitude can be achieved by an additional confinement in the spatial domain.
To this end we design an assembly of a tapered $\lambda$/2 CPW resonator and a patch antenna separated by a dielectric (see figure~\ref{fig:1}C and figure~\ref{fig:2}B).
The patch antenna is facing the rectangular waveguide.
The whole assembly is coupled capacitively to the TE$_{10}$ mode of the rectangular waveguide.
The resonant mode of the assembly in which we are interested can be described as follows.
The tapered resonator part is excited strongly, with mirror charges and corresponding current densities forming on the patch antenna.
Additional oscillating current densities are forming in the part of the ground plane surrounding the tapered resonator.
Consequently, the $B_1$ field is mainly concentrated inside the dielectric and oriented perpendicular to both the bias field and the axis of the tapered resonator.
A considerable field also exists outside the dielectric above the center region of the tapered resonator.
Hence, the perfect position for an NV center is inside the dielectric, which would then be the diamond itself.
Another alternative is placing a diamond on top of the resonator and using an NV center close to the center of the tapered resonator.
%If the patch antenna is omitted, the coupling of the tapered CPW resonator to the TE$_{10}$ mode of the rectangular waveguide is drastically reduced.}
%\out{
%This is implemented by a tapered $\lambda$/2 coplanar waveguide (CPW) resonator.
%Coupling from the waveguide to the resonator mode is \out{efficiently} performed \add{efficiently} by an intermediate copper plate.
%Both structures are fabricated on a dielectric (see figure \ref{fig:1}).
%\\
%\indent
%Since the waveguide mode is TE$_{01}$, an \out{AC}\add{alternating} current in transversal direction is induced in the rectangular copper plate.
%\out{This generates electric fields in longitudinal direction thus exciting}\add{The generated longitudinal electric field excites} the resonator.
%\add{Above and below} the CPW resonator (i.e. outside and inside the dielectric respectively) the $B_{1}$ field has the required polarization.
%}
\\
\indent
Because of the commercial availability of dielectric media, the thickness was fixed to $100\,\mu$m in our experiment.
We do not have sufficiently large diamond samples to use diamond as the dielectric material.
The strength of the $B_1$ field inside the dielectric would increase for a thinner material.
%\out{The strength of the capacitive coupling can be controlled by the thickness of the dielectric, where smaller thicknesses lead to larger coupling.
%The optimum thickness predicted by the simulations is x mm.}
However, critical coupling to the rectangular waveguide can be achieved in any case by adjusting geometric parameters of the CPW resonator (see table~\ref{tab:resonators}).
%\out{Critical coupling could hence
%\comment{MP: The dependance of coupling type and dielectric thickness has not been discussed in the text}
%\comment{PN: You haven't checked whether critical coupling was achieved, have you?}
%not be achieved, but by shaping the geometry of the resonator and \out{of} the intermediate metal plate, the coupling strength was improved.}
\\
\indent
By narrowing the CPW resonator in the center, the current density and, accordingly, the $B_1$ field strength close to the waist are enhanced (see figure \ref{fig:2}B).
%\comment{PN: passt hier nicht! ``The surrounding ground plate suppresses radiative losses.''}
We have simulated and produced two CPW resonators with mainly two different widths of the waist (see table~\ref{tab:resonators}).
For a resonator tapered to a waist of $3\,\mu$m width (resonator \#2),
%\out{a conversion factor of 35.1 G/$\sqrt{W}$ (conversion in Rabi frequency 49.3 MHz/$\sqrt{W}$) is predicted by simulations 20 $\mu$m above the resonator.}
the simulated MW efficiency $20\,\mu$m above the resonator amounts to $C_{\text{mag}}=5.3\,\mathrm{mT/\sqrt{W}}$ ($C_{\text{Rabi}}=105.0\,\mathrm{MHz/\sqrt{W}}$).
\\
\indent
The ground plane around the $\lambda/2$ resonator reduces radiative losses and the patch antenna leads to a concentration of MW energy inside the resonator assembly.
Dielectric materials between CPW resonator and patch antenna with low loss and high dielectric constant $\varepsilon$ can mitigate effects of dielectrics outside of the resonator assembly.
Apart from losses, dielectrics such as diamond or immersion oil on the resonator assembly also lead to a shift of the resonance (see figure~\ref{fig:3}D).
Conductive losses can be reduced by smooth metal surfaces.
For CPW resonator \#2 we have simulated a quality factor of $Q_{\text{2,sim}}=89$, where the contributions of conductive, dielectric and radiative losses are estimated to be 192, 891 and 204, respectively.
%A trade off has to be achieved for the waist region, where smaller width leads to higher fields but also to higher conductive losses.
%\out{\\
%\indent
%Figure \ref{fig:2} B shows \out{a side view of} the $H_{1}$ field.
%The dielectric could be the diamond itself.
%In this case optical access is only feasible for NV centers located in the gap of the CPW resonator structure\add{, which results in a misalignment of the $H_{1}$ field, reducing the effective magnetic field component perpendicular to the spin quantization axis.}
%%\out{It has to be considered that there the x-component of the H field decreases.}
%On the other hand, the field is more concentrated inside the dielectric than above the CPW resonator.
%%Nevertheless for an available diamond membrane of 70 $\mu$m thickness a conversion factor of x G/$\sqrt{W}$ (conversion in Rabi frequency x MHz/$\sqrt{W}$) can be expected in the gap 20 $\mu$m away from the CPW resonator in y direction and 20 $\mu$m deep in the diamond.
%%The smaller distance between patch antenna and CPW resonator leads to an improved coupling yielding higher conversion factors 20 $\mu$m above the CPW resonator: x G/$\sqrt{W}$ (in Rabi frequency x MHz/$\sqrt{W}$).
%}
\\
\indent
We have to consider additional resonances of the whole assembly, due to similar length scales for all coupled parts.
In fact, the patch antenna utilized for improved field concentration can have a resonance close to that of the dipole antenna.
For improper dimensions of the assembly the patch antenna almost completely cancels the transverse $B_1$ field above and below the CPW resonator.
%\out{The intermediate copper plate can also have a resonance.
%Since the plate and the ground of the CPW resonator build together the resonant patch antenna a disruption at the gaps of the CPW resonator leads to an additional node in the center of the field pattern.}\comment{MP: I don't understand anything.}
%\out{Hence in the center if both are on resonance the patch antenna and the CPW resonator are oscillating out-of-phase to each other resulting in decreased B$_{1}$ field strength.}
\\
\indent
In addition to electron spin manipulation by the CPW resonator, a wire or a coil can deliver radio frequencies in order to excite nuclear spin transitions (see figure \ref{fig:4}B).
\\
\indent
The CPW resonators were fabricated out of double sided printed circuit boards (Rogers Ultralam 3850, see table~\ref{tab:resonators}) by optical lithography and wet chemical etching.

\section{Results and Discussion}

\subsection{Single spin optically detected magnetic resonance}
\begin{figure}[t]
\includegraphics[width=\columnwidth]{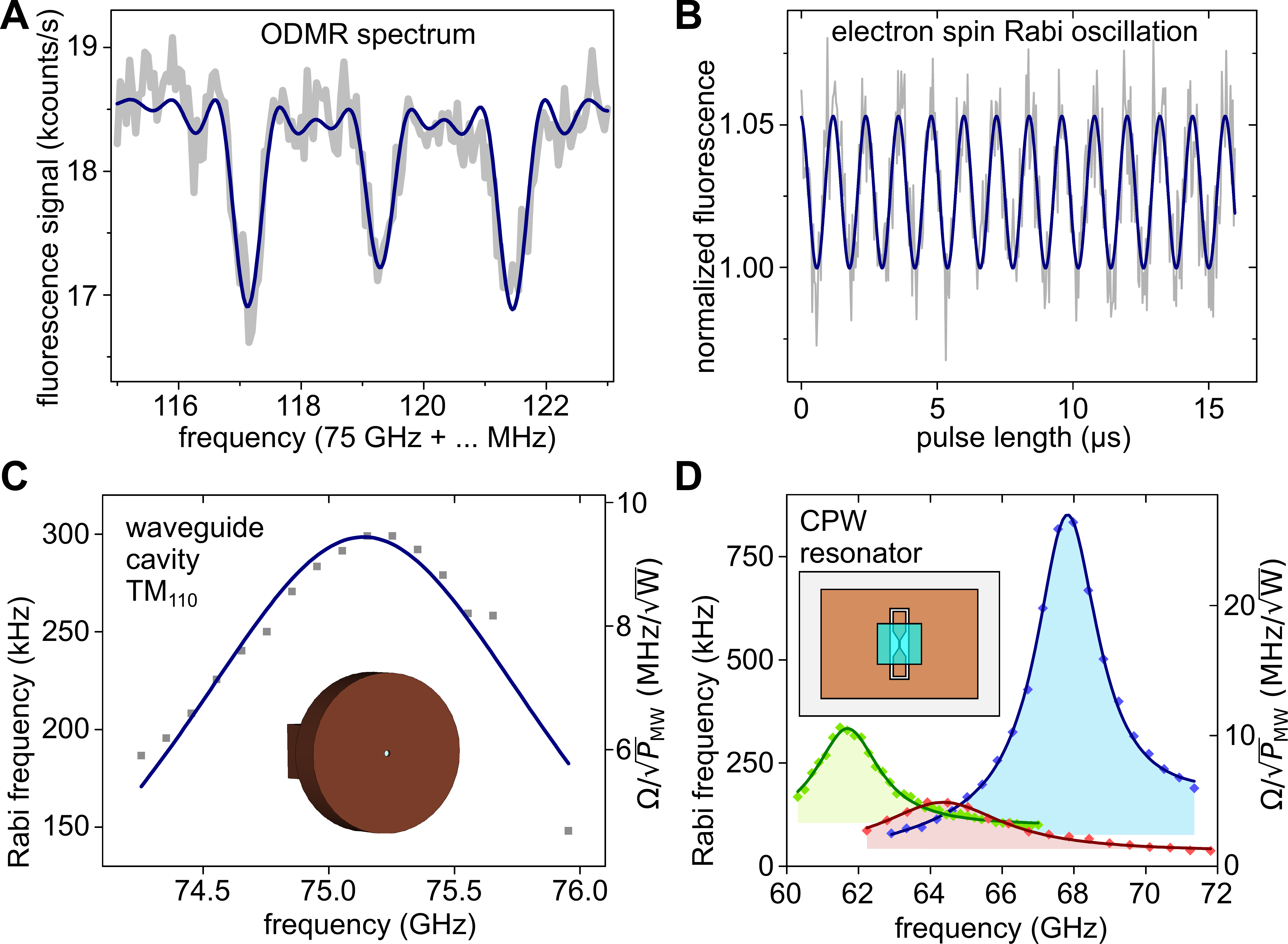}
\caption
	{	\textbf{Spin manipulation at high microwave frequencies.}
		\textbf{A,} ODMR spectrum of the $m_{S}=0$ to $m_{S}=-1$ transition at a magnetic field of $2.78\,$T.
		The $2.16\,$MHz hyperfine splitting caused by the host $^{14}\text{N}$ nuclear spin is clearly resolved.
		\textbf{B,} NV electron spin Rabi oscillations between the $m_{S}=0$ and the $m_{S}=-1$ spin state.
		A sinusoidal fit reveals a Rabi frequency of $906\,$kHz.
		For \textbf{C,} the Rabi frequency depending on the microwave transition frequency was measured for the cavity resonator.
		The magnetic field was swept to shift the spin transition to the respective frequency.
		The plotted Rabi frequency is a measure for the $B_1$ field of the MW.
		The same was done in \textbf{D,} for different configurations of CPW resonators:
		%The blue resonance was measured for a waist of \SI{3}{\micro\meter}, the green and red for a waist of \SI{80}{\micro\meter} (See resonators \#1 and \#2 in table~\ref{tab:resonators} respectively).
		The green and red resonances are measured for resonator \#1 and the blue resonance was measured for resonator \#2.
		For the blue and the green resonance, the diamond was centered above the antenna, while it was shifted away from the waist for the red curve.
		The different amplitudes are caused by differing NV center positions with respect to the waist region.
		%\out{(blue) waist $10\,\mu$m, NV laterally close to waist, $20\,\mu$m above waist.
		%(green) waist $80\,\mu$m, NV laterally not at waist, $20\,\mu$m above waist.
		%(red) same as green but diamond shifted away from waist part.}
		\label{fig:3}
}
\end{figure}

The introduced MW resonators enabled the implementation of single spin ODMR at high magnetic fields.
The basic sequence for a typical pulsed ODMR measurement consists of a spin state reading and polarizing laser pulse accompanied by fluorescence detection, followed by a MW pulse (compare figure~\ref{fig:4}C).
This basic sequence is continuously applied \cite{steiner_universal_2010}.
%\out{The resulting spin state is then read out by another short laser pulse.}
If the MW frequency corresponds to the spin transition frequency, less fluorescence is emitted.
In figure~\ref{fig:3}A, an ODMR spectrum of the $m_{S}=0$ to $m_{S}=-1$ transition is visible.
Due to the hyperfine interaction with the NV center's \textsuperscript{14}N nuclear spin, the transition is split into three distinctive features.
The applied magnetic field is about $2.78\,$T.
\\
\indent
Varying the length of the microwave pulse results in coherent oscillations of the spin projection in z direction (see figure~\ref{fig:3}B).
Since the frequency of MW induced Rabi oscillations is proportional to $B_1$, measuring it as a function of the microwave transition frequency (and thus the external magnetic field) provides information about the performance of the microwave assembly.
The results for the circular waveguide cavity (TM$_{110}$ mode) can be seen in figure~\ref{fig:3}B.
The NV center was located in the center of the optical hole and $\approx 20\,\mu$m deep in the diamond.
The measured quality factor is $Q_{\text{c}}=56$.
For a maximum Rabi frequency of $\Omega=300\,$kHz, the MW efficiency is $C_{\text{Rabi}}=0.75\,\mathrm{MHz/\sqrt{W}}$.
This factor is well below the theoretically predicted one.
In the simulations, the conductive loss caused by the imperfect connection between the cavity element and the metal termination plates was neglected.
Moreover, the surface roughness of the copper film containing the coupling hole can be improved.
As discussed above, a cavity made out of polished diamond would mitigate these loss mechanisms and enhance the concentration of the MW field because of a smaller mode volume.
In the experiment, the diameter of the optical hole was $\approx 500\,\mu$m to ensure a large area of detectable NVs.
This can be scaled down in order to reduce the radiative losses.
\\
\indent
We have produced two CPW resonators mainly differing by the width (see table~\ref{tab:resonators}) of their waists while all other parameters where optimized for strongest $B_1$ field $\approx 20\,\mu$m above the waist.
The parameters given in the table have been confirmed by optical microscopy.
CPW resonator \#1 has a large waist of $85\,\mu$m compared to resonator \#2 with a waist of $3\,\mu$m.
Hence, resonator \#2 is expected to show a larger $B_1$ field close to the waist.
\\
\indent
%In the case of the CPW resonator, the resonance was measured using an NV center $\approx$ 20 $\mu$m above the resonator.
The resonance curves in figure~\ref{fig:3}D were again recorded by measuring field (and therefore MW frequency) dependent Rabi oscillations.
The different colors correspond to different configurations of diamond and resonators.
The green and the red curves are measured using resonator \#1.
In the case of the green curve, the $1\,\text{mm}\times1\,$mm diamond is fully covering the CPW resonator while in the case of the red plot the diamond was covering half of the resonator.
We attribute the $2.66\,$GHz shift of the resonance frequency to the influence of the diamond and its large dielectric constant as is predicted by the simulation results (see figure~\ref{fig:2}B).
The differences in amplitude are mainly due to different distances of the measured NVs to the waist of the resonator.
The maximum MW efficiency measured for this resonator is $C_{\text{Rabi}}=10.6\,\mathrm{MHz/\sqrt{W}}$ (green curve) for an NV center $20\,\mu$m above the waist. 
\\
\indent
The blue curve belongs to resonator \#2 with the diamond centered on the resonator.
%In the case of the $85\,\mu\text{m}$ waist CPW resonator the achieved Rabi conversion is $10.6\,\text{MHz}/\sqrt{\text{W}}$ (green curve).
The MW efficiency reaches $C_{\text{Rabi}}=27\,\mathrm{MHz/\sqrt{W}}$ for an NV center $20\,\mu$m above the resonator close to the waist where the tapering has a width of $20\,\mu$m.
At the position of the waist (with potentially even higher MW efficiency) high background fluorescence from the dielectric material prevented ODMR measurements.
In conclusion, decreasing the waist leads to a further spatial concentration and thus enhancement of the $B_{1}$ field.
\\
\indent
The quality factors of CPW resonators \#1 and \#2 are $Q_1=39$ and $Q_2=48$ respectively, whereas the waveguide cavity reaches $Q_{\text{c}}=56$.
Both waveguide cavity and CPW resonator are useful for our experiments.
Their linewidths respectively quality factors, however, can still be improved.
The advantage of the CPW resonator approach is the concurrent spectral and spatial confinement yielding higher MW efficiencies.

\subsection{Quantum nondemolition measurements and coherent manipulation of single nuclear spin}
\begin{figure}[t]
\includegraphics[width=\columnwidth]{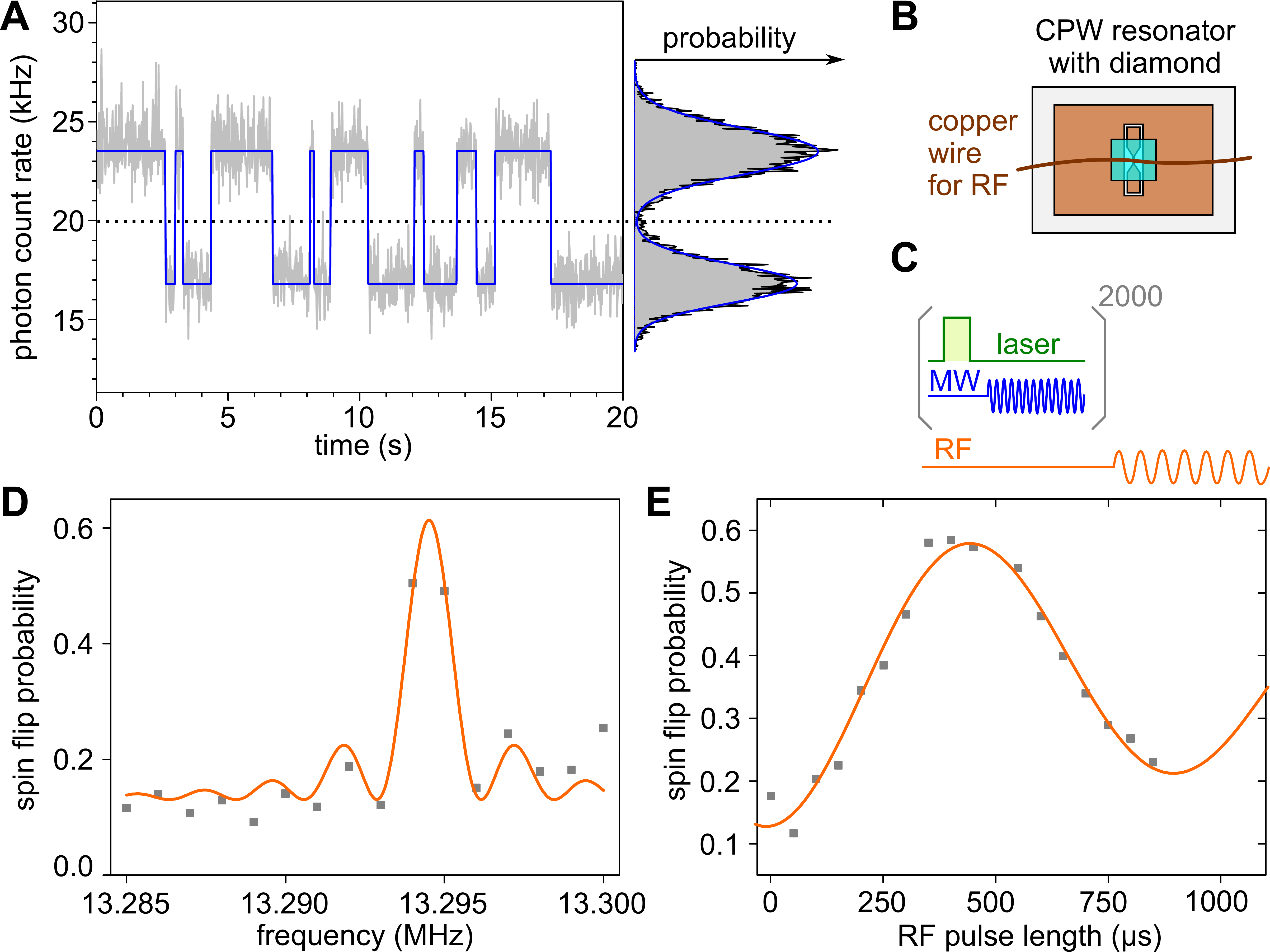}
\caption
	{	\textbf{Single-shot readout and NMR of single nitrogen nuclear spin.}
		\textbf{A,} Continuous readout of the nitrogen nuclear spin reveals quantum jumps.
		The high fluorescence level corresponds to the $m_{I}=-1,0$ nuclear spin states, the low fluorescence level to $m_{I}=+1$.
		The blue line is the result of a two-state Hidden-Markov-Model analysis of the fluorescence timetrace \cite{neumann_single-shot_2010}.
		The fluorescence photon counting histogram is shown on the right.
	\textbf{B,} For coherent nuclear spin control, a copper wire was placed across the diamond mounted on top the coplanar waveguide resonator.
	\textbf{C,} Single shot readout scheme realized by repetitive (2000) nuclear spin state selective flips of the electron spin (MW pulses) and subsequent readout (laser pulse).
	In between readout steps the nuclear spin is coherently manipulated via resonant RF fields.
	For \textbf{D,} the spin flip probability was measured depending on the frequency of an applied RF field, resulting in a maximum at the nuclear spin resonance.
	In \textbf{E,} the microwave pulse length at the resonance frequency was varied, leading to coherent spin oscillations.
		\label{fig:4}
	}
\end{figure}

As explained in the introduction, one of the main reasons to investigate the behavior of NV centers in high magnetic field is the control of surrounding nuclear spins.
Readout of these proximal spins is facilitated by mapping the population onto the NV electron spin by applying a $\text{C}_{\text{n}}\text{NOT}_{\text{e}}$-gate, and consecutive electron spin readout \cite{neumann_single-shot_2010}.
The CNOT-gate is implemented by a nuclear spin state selective electron $\pi$-pulse (see figure~\ref{fig:4}C).
In order to be able to read out the nuclear spin state in a single shot, this measurement has to be repeated on the order of 1000 times, without the nuclear spin to flip.
As has been shown previously, the main source of longitudinal relaxation for the nitrogen nuclear spin are flip-flop events with the corresponding electron spin.
The probability for this to occur can be reduced quadratically with the applied magnetic field.
\\
\indent
Exemplary continuous single shot measurements of the \textsuperscript{14}N nuclear spin state can be seen in figure~\ref{fig:4}A.
The measurement sequence is the same as for pulsed ODMR, but with a fixed microwave frequency.
The fluorescence timetrace reveals quantum jumps between two distinct levels, corresponding to different nuclear spin states (depicted by the blue state trace).
On the right side, a histogram of the fluorescence timetrace shows two separable poissonian distributions (the blue line depicts a fit of the superposition of two poissonians).
This enables single shot readout, as well as initialization by measurement of the nitrogen nuclear spin.
The fluorescence timetrace shows longitudinal relaxation times of the \textsuperscript{14}N nuclear spin on the order of seconds (see figure~\ref{fig:4}A) during continuous projective measurements, an improvement of one order compared to previous measurements performed at $\approx 0.65\,$T \cite{neumann_single-shot_2010}.
\\
\indent
To demonstrate coherent control of the nuclear spin, the NMR spectrum and Rabi oscillations are measured.
A $50\,\mu$m wire is placed across the diamond surface and connected to a RF source to enable nuclear spin manipulation at transition frequencies of several MHz (see figure~\ref{fig:4}B).
\\
\indent
The nuclear spin measurements follow a common approach:
The nuclear spin is measured two times in a single shot, both measurements are then correlated to see if a spin flip occurred.
This probability is measured for different kinds of spin manipulation in between the two nuclear spin state measurements.
By sweeping the frequency of the applied microwave, an NMR spectrum can be recorded (see figure~\ref{fig:4}D).
Varying the microwave pulse length at the resonant frequency leads to coherent Rabi oscillations (see figure~\ref{fig:4}E).

\subsection{Conclusion}
In this study, electron spin resonance and optical microscopy in the E-band is shown to be feasible by exploiting MW resonators.
Spectral confinement is utilized in the cylindrical MW cavity approach.
The waveguide to tapered CPW resonator allows additional confinement in the spatial domain.
A high MW efficiency of $27.0\,\mathrm{MHz/\sqrt{W}}$ ($1.36\,\mathrm{mT/ \sqrt{W}}$) enables ODMR measurements even with a low-power MW source.
At a magnetic field of $2.78\,$T single-shot readout of single nuclear spin reveals seconds-long longitudinal relaxation times.
In addition to readout, coherent control of the nuclear spin via an additional copper wire is shown.
\\
\indent
In principle the presented CPW resonator approach is extendable to even higher frequencies and corresponding magnetic fields.
Then as well as in the current frequency range, free space optics might be used for focusing millimeter waves onto a suitable antenna assembly incorporating single optically addressable spins.
While preparing the manuscript we became aware of a demonstration of ODMR of single electron spins at $4.2\,$T \cite{stepanov_high-frequency_2015} where our approach might also lead to improvements.

\begin{acknowledgments}
We thank A. Denisenko, A. Momenzadeh and F. F\'avaro de Oliveira for sample preparation.
Furthermore, we thank G. Untereiner for TM cavity production.
We acknowledge financial support by the German Science Foundation (SFB-TR 21, SPP1601), the EU (SIQS) and the JST.
\end{acknowledgments}

\bibliography{RefsZotero}

\begin{thebibliography}{31}%
\makeatletter
\providecommand \@ifxundefined [1]{%
 \@ifx{#1\undefined}
}%
\providecommand \@ifnum [1]{%
 \ifnum #1\expandafter \@firstoftwo
 \else \expandafter \@secondoftwo
 \fi
}%
\providecommand \@ifx [1]{%
 \ifx #1\expandafter \@firstoftwo
 \else \expandafter \@secondoftwo
 \fi
}%
\providecommand \natexlab [1]{#1}%
\providecommand \enquote  [1]{``#1''}%
\providecommand \bibnamefont  [1]{#1}%
\providecommand \bibfnamefont [1]{#1}%
\providecommand \citenamefont [1]{#1}%
\providecommand \href@noop [0]{\@secondoftwo}%
\providecommand \href [0]{\begingroup \@sanitize@url \@href}%
\providecommand \@href[1]{\@@startlink{#1}\@@href}%
\providecommand \@@href[1]{\endgroup#1\@@endlink}%
\providecommand \@sanitize@url [0]{\catcode `\\12\catcode `\$12\catcode
  `\&12\catcode `\#12\catcode `\^12\catcode `\_12\catcode `\%12\relax}%
\providecommand \@@startlink[1]{}%
\providecommand \@@endlink[0]{}%
\providecommand \url  [0]{\begingroup\@sanitize@url \@url }%
\providecommand \@url [1]{\endgroup\@href {#1}{\urlprefix }}%
\providecommand \urlprefix  [0]{URL }%
\providecommand \Eprint [0]{\href }%
\providecommand \doibase [0]{http://dx.doi.org/}%
\providecommand \selectlanguage [0]{\@gobble}%
\providecommand \bibinfo  [0]{\@secondoftwo}%
\providecommand \bibfield  [0]{\@secondoftwo}%
\providecommand \translation [1]{[#1]}%
\providecommand \BibitemOpen [0]{}%
\providecommand \bibitemStop [0]{}%
\providecommand \bibitemNoStop [0]{.\EOS\space}%
\providecommand \EOS [0]{\spacefactor3000\relax}%
\providecommand \BibitemShut  [1]{\csname bibitem#1\endcsname}%
\let\auto@bib@innerbib\@empty
%</preamble>
\bibitem [{\citenamefont {Wrachtrup}\ \emph
  {et~al.}(1993{\natexlab{a}})\citenamefont {Wrachtrup}, \citenamefont {von
  Borczyskowski}, \citenamefont {Bernard}, \citenamefont {Orritt},\ and\
  \citenamefont {Brown}}]{wrachtrup_optical_1993}%
  \BibitemOpen
  \bibfield  {author} {\bibinfo {author} {\bibfnamefont {J.}~\bibnamefont
  {Wrachtrup}}, \bibinfo {author} {\bibfnamefont {C.}~\bibnamefont {von
  Borczyskowski}}, \bibinfo {author} {\bibfnamefont {J.}~\bibnamefont
  {Bernard}}, \bibinfo {author} {\bibfnamefont {M.}~\bibnamefont {Orritt}}, \
  and\ \bibinfo {author} {\bibfnamefont {R.}~\bibnamefont {Brown}},\ }\href
  {\doibase 10.1038/363244a0} {\bibfield  {journal} {\bibinfo  {journal}
  {Nature}\ }\textbf {\bibinfo {volume} {363}},\ \bibinfo {pages} {244}
  (\bibinfo {year} {1993}{\natexlab{a}})}\BibitemShut {NoStop}%
\bibitem [{\citenamefont {Wrachtrup}\ \emph
  {et~al.}(1993{\natexlab{b}})\citenamefont {Wrachtrup}, \citenamefont {von
  Borczyskowski}, \citenamefont {Bernard}, \citenamefont {Orrit},\ and\
  \citenamefont {Brown}}]{wrachtrup_optically_1993}%
  \BibitemOpen
  \bibfield  {author} {\bibinfo {author} {\bibfnamefont {J.}~\bibnamefont
  {Wrachtrup}}, \bibinfo {author} {\bibfnamefont {C.}~\bibnamefont {von
  Borczyskowski}}, \bibinfo {author} {\bibfnamefont {J.}~\bibnamefont
  {Bernard}}, \bibinfo {author} {\bibfnamefont {M.}~\bibnamefont {Orrit}}, \
  and\ \bibinfo {author} {\bibfnamefont {R.}~\bibnamefont {Brown}},\ }\href
  {\doibase 10.1103/PhysRevLett.71.3565} {\bibfield  {journal} {\bibinfo
  {journal} {Physical Review Letters}\ }\textbf {\bibinfo {volume} {71}},\
  \bibinfo {pages} {3565} (\bibinfo {year} {1993}{\natexlab{b}})}\BibitemShut
  {NoStop}%
\bibitem [{\citenamefont {Jelezko}\ \emph {et~al.}(2004)\citenamefont
  {Jelezko}, \citenamefont {Gaebel}, \citenamefont {Popa}, \citenamefont
  {Domhan}, \citenamefont {Gruber},\ and\ \citenamefont
  {Wrachtrup}}]{jelezko_observation_2004}%
  \BibitemOpen
  \bibfield  {author} {\bibinfo {author} {\bibfnamefont {F.}~\bibnamefont
  {Jelezko}}, \bibinfo {author} {\bibfnamefont {T.}~\bibnamefont {Gaebel}},
  \bibinfo {author} {\bibfnamefont {I.}~\bibnamefont {Popa}}, \bibinfo {author}
  {\bibfnamefont {M.}~\bibnamefont {Domhan}}, \bibinfo {author} {\bibfnamefont
  {A.}~\bibnamefont {Gruber}}, \ and\ \bibinfo {author} {\bibfnamefont
  {J.}~\bibnamefont {Wrachtrup}},\ }\href {\doibase
  10.1103/PhysRevLett.93.130501} {\bibfield  {journal} {\bibinfo  {journal}
  {Physical Review Letters}\ }\textbf {\bibinfo {volume} {93}},\ \bibinfo
  {pages} {130501} (\bibinfo {year} {2004})}\BibitemShut {NoStop}%
\bibitem [{\citenamefont {Weber}\ \emph {et~al.}(2011)\citenamefont {Weber},
  \citenamefont {Koehl}, \citenamefont {Varley}, \citenamefont {Janotti},
  \citenamefont {Buckley}, \citenamefont {Walle},\ and\ \citenamefont
  {Awschalom}}]{weber_defects_2011}%
  \BibitemOpen
  \bibfield  {author} {\bibinfo {author} {\bibfnamefont {J.~R.}\ \bibnamefont
  {Weber}}, \bibinfo {author} {\bibfnamefont {W.~F.}\ \bibnamefont {Koehl}},
  \bibinfo {author} {\bibfnamefont {J.~B.}\ \bibnamefont {Varley}}, \bibinfo
  {author} {\bibfnamefont {A.}~\bibnamefont {Janotti}}, \bibinfo {author}
  {\bibfnamefont {B.~B.}\ \bibnamefont {Buckley}}, \bibinfo {author}
  {\bibfnamefont {C.~G. V.~d.}\ \bibnamefont {Walle}}, \ and\ \bibinfo {author}
  {\bibfnamefont {D.~D.}\ \bibnamefont {Awschalom}},\ }\href {\doibase
  10.1063/1.3578264} {\bibfield  {journal} {\bibinfo  {journal} {Applied
  Physics Letters}\ }\textbf {\bibinfo {volume} {109}},\ \bibinfo {pages}
  {102417} (\bibinfo {year} {2011})}\BibitemShut {NoStop}%
\bibitem [{\citenamefont {Pla}\ \emph {et~al.}(2013)\citenamefont {Pla},
  \citenamefont {Tan}, \citenamefont {Dehollain}, \citenamefont {Lim},
  \citenamefont {Morton}, \citenamefont {Zwanenburg}, \citenamefont {Jamieson},
  \citenamefont {Dzurak},\ and\ \citenamefont
  {Morello}}]{pla_high-fidelity_2013}%
  \BibitemOpen
  \bibfield  {author} {\bibinfo {author} {\bibfnamefont {J.~J.}\ \bibnamefont
  {Pla}}, \bibinfo {author} {\bibfnamefont {K.~Y.}\ \bibnamefont {Tan}},
  \bibinfo {author} {\bibfnamefont {J.~P.}\ \bibnamefont {Dehollain}}, \bibinfo
  {author} {\bibfnamefont {W.~H.}\ \bibnamefont {Lim}}, \bibinfo {author}
  {\bibfnamefont {J.~J.~L.}\ \bibnamefont {Morton}}, \bibinfo {author}
  {\bibfnamefont {F.~A.}\ \bibnamefont {Zwanenburg}}, \bibinfo {author}
  {\bibfnamefont {D.~N.}\ \bibnamefont {Jamieson}}, \bibinfo {author}
  {\bibfnamefont {A.~S.}\ \bibnamefont {Dzurak}}, \ and\ \bibinfo {author}
  {\bibfnamefont {A.}~\bibnamefont {Morello}},\ }\href {\doibase
  10.1038/nature12011} {\bibfield  {journal} {\bibinfo  {journal} {Nature}\
  }\textbf {\bibinfo {volume} {496}},\ \bibinfo {pages} {334} (\bibinfo {year}
  {2013})}\BibitemShut {NoStop}%
\bibitem [{\citenamefont {Siyushev}\ \emph {et~al.}(2014)\citenamefont
  {Siyushev}, \citenamefont {Xia}, \citenamefont {Reuter}, \citenamefont
  {Jamali}, \citenamefont {Zhao}, \citenamefont {Yang}, \citenamefont {Duan},
  \citenamefont {Kukharchyk}, \citenamefont {Wieck}, \citenamefont {Kolesov},\
  and\ \citenamefont {Wrachtrup}}]{siyushev_coherent_2014}%
  \BibitemOpen
  \bibfield  {author} {\bibinfo {author} {\bibfnamefont {P.}~\bibnamefont
  {Siyushev}}, \bibinfo {author} {\bibfnamefont {K.}~\bibnamefont {Xia}},
  \bibinfo {author} {\bibfnamefont {R.}~\bibnamefont {Reuter}}, \bibinfo
  {author} {\bibfnamefont {M.}~\bibnamefont {Jamali}}, \bibinfo {author}
  {\bibfnamefont {N.}~\bibnamefont {Zhao}}, \bibinfo {author} {\bibfnamefont
  {N.}~\bibnamefont {Yang}}, \bibinfo {author} {\bibfnamefont {C.}~\bibnamefont
  {Duan}}, \bibinfo {author} {\bibfnamefont {N.}~\bibnamefont {Kukharchyk}},
  \bibinfo {author} {\bibfnamefont {A.~D.}\ \bibnamefont {Wieck}}, \bibinfo
  {author} {\bibfnamefont {R.}~\bibnamefont {Kolesov}}, \ and\ \bibinfo
  {author} {\bibfnamefont {J.}~\bibnamefont {Wrachtrup}},\ }\href {\doibase
  10.1038/ncomms4895} {\bibfield  {journal} {\bibinfo  {journal} {Nature
  Communications}\ }\textbf {\bibinfo {volume} {5}} (\bibinfo {year} {2014}),\
  10.1038/ncomms4895},\ \bibinfo {note} {arXiv:1405.5258 [cond-mat,
  physics:quant-ph]}\BibitemShut {NoStop}%
\bibitem [{\citenamefont {Toyli}\ \emph {et~al.}(2010)\citenamefont {Toyli},
  \citenamefont {Weis}, \citenamefont {Fuchs}, \citenamefont {Schenkel},\ and\
  \citenamefont {Awschalom}}]{toyli_chip-scale_2010}%
  \BibitemOpen
  \bibfield  {author} {\bibinfo {author} {\bibfnamefont {D.~M.}\ \bibnamefont
  {Toyli}}, \bibinfo {author} {\bibfnamefont {C.~D.}\ \bibnamefont {Weis}},
  \bibinfo {author} {\bibfnamefont {G.~D.}\ \bibnamefont {Fuchs}}, \bibinfo
  {author} {\bibfnamefont {T.}~\bibnamefont {Schenkel}}, \ and\ \bibinfo
  {author} {\bibfnamefont {D.~D.}\ \bibnamefont {Awschalom}},\ }\href {\doibase
  10.1021/nl102066q} {\bibfield  {journal} {\bibinfo  {journal} {Nano Letters}\
  }\textbf {\bibinfo {volume} {10}},\ \bibinfo {pages} {3168} (\bibinfo {year}
  {2010})}\BibitemShut {NoStop}%
\bibitem [{\citenamefont {Morello}\ \emph {et~al.}(2010)\citenamefont
  {Morello}, \citenamefont {Pla}, \citenamefont {Zwanenburg}, \citenamefont
  {Chan}, \citenamefont {Tan}, \citenamefont {Huebl}, \citenamefont {Mottonen},
  \citenamefont {Nugroho}, \citenamefont {Yang}, \citenamefont {van Donkelaar},
  \citenamefont {Alves}, \citenamefont {Jamieson}, \citenamefont {Escott},
  \citenamefont {Hollenberg}, \citenamefont {Clark},\ and\ \citenamefont
  {Dzurak}}]{morello_single-shot_2010}%
  \BibitemOpen
  \bibfield  {author} {\bibinfo {author} {\bibfnamefont {A.}~\bibnamefont
  {Morello}}, \bibinfo {author} {\bibfnamefont {J.~J.}\ \bibnamefont {Pla}},
  \bibinfo {author} {\bibfnamefont {F.~A.}\ \bibnamefont {Zwanenburg}},
  \bibinfo {author} {\bibfnamefont {K.~W.}\ \bibnamefont {Chan}}, \bibinfo
  {author} {\bibfnamefont {K.~Y.}\ \bibnamefont {Tan}}, \bibinfo {author}
  {\bibfnamefont {H.}~\bibnamefont {Huebl}}, \bibinfo {author} {\bibfnamefont
  {M.}~\bibnamefont {Mottonen}}, \bibinfo {author} {\bibfnamefont {C.~D.}\
  \bibnamefont {Nugroho}}, \bibinfo {author} {\bibfnamefont {C.}~\bibnamefont
  {Yang}}, \bibinfo {author} {\bibfnamefont {J.~A.}\ \bibnamefont {van
  Donkelaar}}, \bibinfo {author} {\bibfnamefont {A.~D.~C.}\ \bibnamefont
  {Alves}}, \bibinfo {author} {\bibfnamefont {D.~N.}\ \bibnamefont {Jamieson}},
  \bibinfo {author} {\bibfnamefont {C.~C.}\ \bibnamefont {Escott}}, \bibinfo
  {author} {\bibfnamefont {L.~C.~L.}\ \bibnamefont {Hollenberg}}, \bibinfo
  {author} {\bibfnamefont {R.~G.}\ \bibnamefont {Clark}}, \ and\ \bibinfo
  {author} {\bibfnamefont {A.~S.}\ \bibnamefont {Dzurak}},\ }\href {\doibase
  10.1038/nature09392} {\bibfield  {journal} {\bibinfo  {journal} {Nature}\
  }\textbf {\bibinfo {volume} {467}},\ \bibinfo {pages} {687} (\bibinfo {year}
  {2010})}\BibitemShut {NoStop}%
\bibitem [{\citenamefont {Kolesov}\ \emph {et~al.}(2013)\citenamefont
  {Kolesov}, \citenamefont {Xia}, \citenamefont {Reuter}, \citenamefont
  {Jamali}, \citenamefont {St{\"o}hr}, \citenamefont {Inal}, \citenamefont
  {Siyushev},\ and\ \citenamefont {Wrachtrup}}]{kolesov_mapping_2013}%
  \BibitemOpen
  \bibfield  {author} {\bibinfo {author} {\bibfnamefont {R.}~\bibnamefont
  {Kolesov}}, \bibinfo {author} {\bibfnamefont {K.}~\bibnamefont {Xia}},
  \bibinfo {author} {\bibfnamefont {R.}~\bibnamefont {Reuter}}, \bibinfo
  {author} {\bibfnamefont {M.}~\bibnamefont {Jamali}}, \bibinfo {author}
  {\bibfnamefont {R.}~\bibnamefont {St{\"o}hr}}, \bibinfo {author}
  {\bibfnamefont {T.}~\bibnamefont {Inal}}, \bibinfo {author} {\bibfnamefont
  {P.}~\bibnamefont {Siyushev}}, \ and\ \bibinfo {author} {\bibfnamefont
  {J.}~\bibnamefont {Wrachtrup}},\ }\href {\doibase
  10.1103/PhysRevLett.111.120502} {\bibfield  {journal} {\bibinfo  {journal}
  {Physical Review Letters}\ }\textbf {\bibinfo {volume} {111}},\ \bibinfo
  {pages} {120502} (\bibinfo {year} {2013})}\BibitemShut {NoStop}%
\bibitem [{\citenamefont {Kolesov}\ \emph {et~al.}(2012)\citenamefont
  {Kolesov}, \citenamefont {Xia}, \citenamefont {Reuter}, \citenamefont
  {St{\"o}hr}, \citenamefont {Zappe}, \citenamefont {Meijer}, \citenamefont
  {Hemmer},\ and\ \citenamefont {Wrachtrup}}]{kolesov_optical_2012}%
  \BibitemOpen
  \bibfield  {author} {\bibinfo {author} {\bibfnamefont {R.}~\bibnamefont
  {Kolesov}}, \bibinfo {author} {\bibfnamefont {K.}~\bibnamefont {Xia}},
  \bibinfo {author} {\bibfnamefont {R.}~\bibnamefont {Reuter}}, \bibinfo
  {author} {\bibfnamefont {R.}~\bibnamefont {St{\"o}hr}}, \bibinfo {author}
  {\bibfnamefont {A.}~\bibnamefont {Zappe}}, \bibinfo {author} {\bibfnamefont
  {J.}~\bibnamefont {Meijer}}, \bibinfo {author} {\bibfnamefont {P.~R.}\
  \bibnamefont {Hemmer}}, \ and\ \bibinfo {author} {\bibfnamefont
  {J.}~\bibnamefont {Wrachtrup}},\ }\href {\doibase 10.1038/ncomms2034}
  {\bibfield  {journal} {\bibinfo  {journal} {Nature Communications}\ }\textbf
  {\bibinfo {volume} {3}},\ \bibinfo {pages} {1029} (\bibinfo {year}
  {2012})}\BibitemShut {NoStop}%
\bibitem [{\citenamefont {Yin}\ \emph {et~al.}(2013)\citenamefont {Yin},
  \citenamefont {Rancic}, \citenamefont {de~Boo}, \citenamefont {Stavrias},
  \citenamefont {McCallum}, \citenamefont {Sellars},\ and\ \citenamefont
  {Rogge}}]{yin_optical_2013}%
  \BibitemOpen
  \bibfield  {author} {\bibinfo {author} {\bibfnamefont {C.}~\bibnamefont
  {Yin}}, \bibinfo {author} {\bibfnamefont {M.}~\bibnamefont {Rancic}},
  \bibinfo {author} {\bibfnamefont {G.~G.}\ \bibnamefont {de~Boo}}, \bibinfo
  {author} {\bibfnamefont {N.}~\bibnamefont {Stavrias}}, \bibinfo {author}
  {\bibfnamefont {J.~C.}\ \bibnamefont {McCallum}}, \bibinfo {author}
  {\bibfnamefont {M.~J.}\ \bibnamefont {Sellars}}, \ and\ \bibinfo {author}
  {\bibfnamefont {S.}~\bibnamefont {Rogge}},\ }\href {\doibase
  10.1038/nature12081} {\bibfield  {journal} {\bibinfo  {journal} {Nature}\
  }\textbf {\bibinfo {volume} {497}},\ \bibinfo {pages} {91} (\bibinfo {year}
  {2013})}\BibitemShut {NoStop}%
\bibitem [{\citenamefont {Waldherr}\ \emph {et~al.}(2014)\citenamefont
  {Waldherr}, \citenamefont {Wang}, \citenamefont {Zaiser}, \citenamefont
  {Jamali}, \citenamefont {Schulte-Herbr{\"u}ggen}, \citenamefont {Abe},
  \citenamefont {Ohshima}, \citenamefont {Isoya}, \citenamefont {Du},
  \citenamefont {Neumann},\ and\ \citenamefont
  {Wrachtrup}}]{waldherr_quantum_2014}%
  \BibitemOpen
  \bibfield  {author} {\bibinfo {author} {\bibfnamefont {G.}~\bibnamefont
  {Waldherr}}, \bibinfo {author} {\bibfnamefont {Y.}~\bibnamefont {Wang}},
  \bibinfo {author} {\bibfnamefont {S.}~\bibnamefont {Zaiser}}, \bibinfo
  {author} {\bibfnamefont {M.}~\bibnamefont {Jamali}}, \bibinfo {author}
  {\bibfnamefont {T.}~\bibnamefont {Schulte-Herbr{\"u}ggen}}, \bibinfo {author}
  {\bibfnamefont {H.}~\bibnamefont {Abe}}, \bibinfo {author} {\bibfnamefont
  {T.}~\bibnamefont {Ohshima}}, \bibinfo {author} {\bibfnamefont
  {J.}~\bibnamefont {Isoya}}, \bibinfo {author} {\bibfnamefont {J.~F.}\
  \bibnamefont {Du}}, \bibinfo {author} {\bibfnamefont {P.}~\bibnamefont
  {Neumann}}, \ and\ \bibinfo {author} {\bibfnamefont {J.}~\bibnamefont
  {Wrachtrup}},\ }\href {\doibase 10.1038/nature12919} {\bibfield  {journal}
  {\bibinfo  {journal} {Nature}\ }\textbf {\bibinfo {volume} {506}},\ \bibinfo
  {pages} {204} (\bibinfo {year} {2014})}\BibitemShut {NoStop}%
\bibitem [{\citenamefont {Dolde}\ \emph {et~al.}(2014)\citenamefont {Dolde},
  \citenamefont {Bergholm}, \citenamefont {Wang}, \citenamefont {Jakobi},
  \citenamefont {Naydenov}, \citenamefont {Pezzagna}, \citenamefont {Meijer},
  \citenamefont {Jelezko}, \citenamefont {Neumann}, \citenamefont
  {Schulte-Herbr{\"u}ggen}, \citenamefont {Biamonte},\ and\ \citenamefont
  {Wrachtrup}}]{dolde_high-fidelity_2014}%
  \BibitemOpen
  \bibfield  {author} {\bibinfo {author} {\bibfnamefont {F.}~\bibnamefont
  {Dolde}}, \bibinfo {author} {\bibfnamefont {V.}~\bibnamefont {Bergholm}},
  \bibinfo {author} {\bibfnamefont {Y.}~\bibnamefont {Wang}}, \bibinfo {author}
  {\bibfnamefont {I.}~\bibnamefont {Jakobi}}, \bibinfo {author} {\bibfnamefont
  {B.}~\bibnamefont {Naydenov}}, \bibinfo {author} {\bibfnamefont
  {S.}~\bibnamefont {Pezzagna}}, \bibinfo {author} {\bibfnamefont
  {J.}~\bibnamefont {Meijer}}, \bibinfo {author} {\bibfnamefont
  {F.}~\bibnamefont {Jelezko}}, \bibinfo {author} {\bibfnamefont
  {P.}~\bibnamefont {Neumann}}, \bibinfo {author} {\bibfnamefont
  {T.}~\bibnamefont {Schulte-Herbr{\"u}ggen}}, \bibinfo {author} {\bibfnamefont
  {J.}~\bibnamefont {Biamonte}}, \ and\ \bibinfo {author} {\bibfnamefont
  {J.}~\bibnamefont {Wrachtrup}},\ }\href {\doibase 10.1038/ncomms4371}
  {\bibfield  {journal} {\bibinfo  {journal} {Nature Communications}\ }\textbf
  {\bibinfo {volume} {5}},\ \bibinfo {pages} {3371} (\bibinfo {year}
  {2014})}\BibitemShut {NoStop}%
\bibitem [{\citenamefont {Togan}\ \emph {et~al.}(2010)\citenamefont {Togan},
  \citenamefont {Chu}, \citenamefont {Trifonov}, \citenamefont {Jiang},
  \citenamefont {Maze}, \citenamefont {Childress}, \citenamefont {Dutt},
  \citenamefont {S{\o}rensen}, \citenamefont {Hemmer}, \citenamefont {Zibrov},\
  and\ \citenamefont {Lukin}}]{togan_quantum_2010}%
  \BibitemOpen
  \bibfield  {author} {\bibinfo {author} {\bibfnamefont {E.}~\bibnamefont
  {Togan}}, \bibinfo {author} {\bibfnamefont {Y.}~\bibnamefont {Chu}}, \bibinfo
  {author} {\bibfnamefont {A.~S.}\ \bibnamefont {Trifonov}}, \bibinfo {author}
  {\bibfnamefont {L.}~\bibnamefont {Jiang}}, \bibinfo {author} {\bibfnamefont
  {J.}~\bibnamefont {Maze}}, \bibinfo {author} {\bibfnamefont {L.}~\bibnamefont
  {Childress}}, \bibinfo {author} {\bibfnamefont {M.~V.~G.}\ \bibnamefont
  {Dutt}}, \bibinfo {author} {\bibfnamefont {A.~S.}\ \bibnamefont
  {S{\o}rensen}}, \bibinfo {author} {\bibfnamefont {P.~R.}\ \bibnamefont
  {Hemmer}}, \bibinfo {author} {\bibfnamefont {A.~S.}\ \bibnamefont {Zibrov}},
  \ and\ \bibinfo {author} {\bibfnamefont {M.~D.}\ \bibnamefont {Lukin}},\
  }\href {\doibase 10.1038/nature09256} {\bibfield  {journal} {\bibinfo
  {journal} {Nature}\ }\textbf {\bibinfo {volume} {466}},\ \bibinfo {pages}
  {730} (\bibinfo {year} {2010})}\BibitemShut {NoStop}%
\bibitem [{\citenamefont {Bernien}\ \emph {et~al.}(2013)\citenamefont
  {Bernien}, \citenamefont {Hensen}, \citenamefont {Pfaff}, \citenamefont
  {Koolstra}, \citenamefont {Blok}, \citenamefont {Robledo}, \citenamefont
  {Taminiau}, \citenamefont {Markham}, \citenamefont {Twitchen}, \citenamefont
  {Childress},\ and\ \citenamefont {Hanson}}]{bernien_heralded_2013-2}%
  \BibitemOpen
  \bibfield  {author} {\bibinfo {author} {\bibfnamefont {H.}~\bibnamefont
  {Bernien}}, \bibinfo {author} {\bibfnamefont {B.}~\bibnamefont {Hensen}},
  \bibinfo {author} {\bibfnamefont {W.}~\bibnamefont {Pfaff}}, \bibinfo
  {author} {\bibfnamefont {G.}~\bibnamefont {Koolstra}}, \bibinfo {author}
  {\bibfnamefont {M.~S.}\ \bibnamefont {Blok}}, \bibinfo {author}
  {\bibfnamefont {L.}~\bibnamefont {Robledo}}, \bibinfo {author} {\bibfnamefont
  {T.~H.}\ \bibnamefont {Taminiau}}, \bibinfo {author} {\bibfnamefont
  {M.}~\bibnamefont {Markham}}, \bibinfo {author} {\bibfnamefont {D.~J.}\
  \bibnamefont {Twitchen}}, \bibinfo {author} {\bibfnamefont {L.}~\bibnamefont
  {Childress}}, \ and\ \bibinfo {author} {\bibfnamefont {R.}~\bibnamefont
  {Hanson}},\ }\href {\doibase 10.1038/nature12016} {\bibfield  {journal}
  {\bibinfo  {journal} {Nature}\ }\textbf {\bibinfo {volume} {497}},\ \bibinfo
  {pages} {86} (\bibinfo {year} {2013})}\BibitemShut {NoStop}%
\bibitem [{\citenamefont {McCamey}\ \emph {et~al.}(2010)\citenamefont
  {McCamey}, \citenamefont {Tol}, \citenamefont {Morley},\ and\ \citenamefont
  {Boehme}}]{mccamey_electronic_2010}%
  \BibitemOpen
  \bibfield  {author} {\bibinfo {author} {\bibfnamefont {D.~R.}\ \bibnamefont
  {McCamey}}, \bibinfo {author} {\bibfnamefont {J.~V.}\ \bibnamefont {Tol}},
  \bibinfo {author} {\bibfnamefont {G.~W.}\ \bibnamefont {Morley}}, \ and\
  \bibinfo {author} {\bibfnamefont {C.}~\bibnamefont {Boehme}},\ }\href
  {\doibase 10.1126/science.1197931} {\bibfield  {journal} {\bibinfo  {journal}
  {Science}\ }\textbf {\bibinfo {volume} {330}},\ \bibinfo {pages} {1652}
  (\bibinfo {year} {2010})}\BibitemShut {NoStop}%
\bibitem [{\citenamefont {Balasubramanian}\ \emph {et~al.}(2008)\citenamefont
  {Balasubramanian}, \citenamefont {Chan}, \citenamefont {Kolesov},
  \citenamefont {Al-Hmoud}, \citenamefont {Tisler}, \citenamefont {Shin},
  \citenamefont {Kim}, \citenamefont {Wojcik}, \citenamefont {Hemmer},
  \citenamefont {Krueger}, \citenamefont {Hanke}, \citenamefont
  {Leitenstorfer}, \citenamefont {Bratschitsch}, \citenamefont {Jelezko},\ and\
  \citenamefont {Wrachtrup}}]{balasubramanian_nanoscale_2008}%
  \BibitemOpen
  \bibfield  {author} {\bibinfo {author} {\bibfnamefont {G.}~\bibnamefont
  {Balasubramanian}}, \bibinfo {author} {\bibfnamefont {I.~Y.}\ \bibnamefont
  {Chan}}, \bibinfo {author} {\bibfnamefont {R.}~\bibnamefont {Kolesov}},
  \bibinfo {author} {\bibfnamefont {M.}~\bibnamefont {Al-Hmoud}}, \bibinfo
  {author} {\bibfnamefont {J.}~\bibnamefont {Tisler}}, \bibinfo {author}
  {\bibfnamefont {C.}~\bibnamefont {Shin}}, \bibinfo {author} {\bibfnamefont
  {C.}~\bibnamefont {Kim}}, \bibinfo {author} {\bibfnamefont {A.}~\bibnamefont
  {Wojcik}}, \bibinfo {author} {\bibfnamefont {P.~R.}\ \bibnamefont {Hemmer}},
  \bibinfo {author} {\bibfnamefont {A.}~\bibnamefont {Krueger}}, \bibinfo
  {author} {\bibfnamefont {T.}~\bibnamefont {Hanke}}, \bibinfo {author}
  {\bibfnamefont {A.}~\bibnamefont {Leitenstorfer}}, \bibinfo {author}
  {\bibfnamefont {R.}~\bibnamefont {Bratschitsch}}, \bibinfo {author}
  {\bibfnamefont {F.}~\bibnamefont {Jelezko}}, \ and\ \bibinfo {author}
  {\bibfnamefont {J.}~\bibnamefont {Wrachtrup}},\ }\href {\doibase
  10.1038/nature07278} {\bibfield  {journal} {\bibinfo  {journal} {Nature}\
  }\textbf {\bibinfo {volume} {455}},\ \bibinfo {pages} {648} (\bibinfo {year}
  {2008})}\BibitemShut {NoStop}%
\bibitem [{\citenamefont {Maze}\ \emph {et~al.}(2008)\citenamefont {Maze},
  \citenamefont {Stanwix}, \citenamefont {Hodges}, \citenamefont {Hong},
  \citenamefont {Taylor}, \citenamefont {Cappellaro}, \citenamefont {Jiang},
  \citenamefont {Dutt}, \citenamefont {Togan}, \citenamefont {Zibrov},
  \citenamefont {Yacoby}, \citenamefont {Walsworth},\ and\ \citenamefont
  {Lukin}}]{maze_nanoscale_2008}%
  \BibitemOpen
  \bibfield  {author} {\bibinfo {author} {\bibfnamefont {J.~R.}\ \bibnamefont
  {Maze}}, \bibinfo {author} {\bibfnamefont {P.~L.}\ \bibnamefont {Stanwix}},
  \bibinfo {author} {\bibfnamefont {J.~S.}\ \bibnamefont {Hodges}}, \bibinfo
  {author} {\bibfnamefont {S.}~\bibnamefont {Hong}}, \bibinfo {author}
  {\bibfnamefont {J.~M.}\ \bibnamefont {Taylor}}, \bibinfo {author}
  {\bibfnamefont {P.}~\bibnamefont {Cappellaro}}, \bibinfo {author}
  {\bibfnamefont {L.}~\bibnamefont {Jiang}}, \bibinfo {author} {\bibfnamefont
  {M.~V.~G.}\ \bibnamefont {Dutt}}, \bibinfo {author} {\bibfnamefont
  {E.}~\bibnamefont {Togan}}, \bibinfo {author} {\bibfnamefont {A.~S.}\
  \bibnamefont {Zibrov}}, \bibinfo {author} {\bibfnamefont {A.}~\bibnamefont
  {Yacoby}}, \bibinfo {author} {\bibfnamefont {R.~L.}\ \bibnamefont
  {Walsworth}}, \ and\ \bibinfo {author} {\bibfnamefont {M.~D.}\ \bibnamefont
  {Lukin}},\ }\href {\doibase 10.1038/nature07279} {\bibfield  {journal}
  {\bibinfo  {journal} {Nature}\ }\textbf {\bibinfo {volume} {455}},\ \bibinfo
  {pages} {644} (\bibinfo {year} {2008})}\BibitemShut {NoStop}%
\bibitem [{\citenamefont {H{\"a}berle}\ \emph {et~al.}(2015)\citenamefont
  {H{\"a}berle}, \citenamefont {Schmid-Lorch}, \citenamefont {Reinhard},\ and\
  \citenamefont {Wrachtrup}}]{haberle_nanoscale_2015-1}%
  \BibitemOpen
  \bibfield  {author} {\bibinfo {author} {\bibfnamefont {T.}~\bibnamefont
  {H{\"a}berle}}, \bibinfo {author} {\bibfnamefont {D.}~\bibnamefont
  {Schmid-Lorch}}, \bibinfo {author} {\bibfnamefont {F.}~\bibnamefont
  {Reinhard}}, \ and\ \bibinfo {author} {\bibfnamefont {J.}~\bibnamefont
  {Wrachtrup}},\ }\href {\doibase 10.1038/nnano.2014.299} {\bibfield  {journal}
  {\bibinfo  {journal} {Nature Nanotechnology}\ }\textbf {\bibinfo {volume}
  {10}},\ \bibinfo {pages} {125} (\bibinfo {year} {2015})}\BibitemShut
  {NoStop}%
\bibitem [{\citenamefont {Mamin}\ \emph {et~al.}(2013)\citenamefont {Mamin},
  \citenamefont {Kim}, \citenamefont {Sherwood}, \citenamefont {Rettner},
  \citenamefont {Ohno}, \citenamefont {Awschalom},\ and\ \citenamefont
  {Rugar}}]{mamin_nanoscale_2013}%
  \BibitemOpen
  \bibfield  {author} {\bibinfo {author} {\bibfnamefont {H.~J.}\ \bibnamefont
  {Mamin}}, \bibinfo {author} {\bibfnamefont {M.}~\bibnamefont {Kim}}, \bibinfo
  {author} {\bibfnamefont {M.~H.}\ \bibnamefont {Sherwood}}, \bibinfo {author}
  {\bibfnamefont {C.~T.}\ \bibnamefont {Rettner}}, \bibinfo {author}
  {\bibfnamefont {K.}~\bibnamefont {Ohno}}, \bibinfo {author} {\bibfnamefont
  {D.~D.}\ \bibnamefont {Awschalom}}, \ and\ \bibinfo {author} {\bibfnamefont
  {D.}~\bibnamefont {Rugar}},\ }\href {\doibase 10.1126/science.1231540}
  {\bibfield  {journal} {\bibinfo  {journal} {Science}\ }\textbf {\bibinfo
  {volume} {339}},\ \bibinfo {pages} {557} (\bibinfo {year}
  {2013})}\BibitemShut {NoStop}%
\bibitem [{\citenamefont {DeVience}\ \emph {et~al.}(2015)\citenamefont
  {DeVience}, \citenamefont {Pham}, \citenamefont {Lovchinsky}, \citenamefont
  {Sushkov}, \citenamefont {Bar-Gill}, \citenamefont {Belthangady},
  \citenamefont {Casola}, \citenamefont {Corbett}, \citenamefont {Zhang},
  \citenamefont {Lukin}, \citenamefont {Park}, \citenamefont {Yacoby},\ and\
  \citenamefont {Walsworth}}]{devience_nanoscale_2015}%
  \BibitemOpen
  \bibfield  {author} {\bibinfo {author} {\bibfnamefont {S.~J.}\ \bibnamefont
  {DeVience}}, \bibinfo {author} {\bibfnamefont {L.~M.}\ \bibnamefont {Pham}},
  \bibinfo {author} {\bibfnamefont {I.}~\bibnamefont {Lovchinsky}}, \bibinfo
  {author} {\bibfnamefont {A.~O.}\ \bibnamefont {Sushkov}}, \bibinfo {author}
  {\bibfnamefont {N.}~\bibnamefont {Bar-Gill}}, \bibinfo {author}
  {\bibfnamefont {C.}~\bibnamefont {Belthangady}}, \bibinfo {author}
  {\bibfnamefont {F.}~\bibnamefont {Casola}}, \bibinfo {author} {\bibfnamefont
  {M.}~\bibnamefont {Corbett}}, \bibinfo {author} {\bibfnamefont
  {H.}~\bibnamefont {Zhang}}, \bibinfo {author} {\bibfnamefont
  {M.}~\bibnamefont {Lukin}}, \bibinfo {author} {\bibfnamefont
  {H.}~\bibnamefont {Park}}, \bibinfo {author} {\bibfnamefont {A.}~\bibnamefont
  {Yacoby}}, \ and\ \bibinfo {author} {\bibfnamefont {R.~L.}\ \bibnamefont
  {Walsworth}},\ }\href {\doibase 10.1038/nnano.2014.313} {\bibfield  {journal}
  {\bibinfo  {journal} {Nature Nanotechnology}\ }\textbf {\bibinfo {volume}
  {10}},\ \bibinfo {pages} {129} (\bibinfo {year} {2015})}\BibitemShut
  {NoStop}%
\bibitem [{\citenamefont {Staudacher}\ \emph {et~al.}(2013)\citenamefont
  {Staudacher}, \citenamefont {Shi}, \citenamefont {Pezzagna}, \citenamefont
  {Meijer}, \citenamefont {Du}, \citenamefont {Meriles}, \citenamefont
  {Reinhard},\ and\ \citenamefont {Wrachtrup}}]{staudacher_nuclear_2013}%
  \BibitemOpen
  \bibfield  {author} {\bibinfo {author} {\bibfnamefont {T.}~\bibnamefont
  {Staudacher}}, \bibinfo {author} {\bibfnamefont {F.}~\bibnamefont {Shi}},
  \bibinfo {author} {\bibfnamefont {S.}~\bibnamefont {Pezzagna}}, \bibinfo
  {author} {\bibfnamefont {J.}~\bibnamefont {Meijer}}, \bibinfo {author}
  {\bibfnamefont {J.}~\bibnamefont {Du}}, \bibinfo {author} {\bibfnamefont
  {C.~A.}\ \bibnamefont {Meriles}}, \bibinfo {author} {\bibfnamefont
  {F.}~\bibnamefont {Reinhard}}, \ and\ \bibinfo {author} {\bibfnamefont
  {J.}~\bibnamefont {Wrachtrup}},\ }\href {\doibase 10.1126/science.1231675}
  {\bibfield  {journal} {\bibinfo  {journal} {Science}\ }\textbf {\bibinfo
  {volume} {339}},\ \bibinfo {pages} {561} (\bibinfo {year}
  {2013})}\BibitemShut {NoStop}%
\bibitem [{\citenamefont {Grotz}\ \emph {et~al.}(2011)\citenamefont {Grotz},
  \citenamefont {Beck}, \citenamefont {Neumann}, \citenamefont {Naydenov},
  \citenamefont {Reuter}, \citenamefont {Reinhard}, \citenamefont {Jelezko},
  \citenamefont {Wrachtrup}, \citenamefont {Schweinfurth}, \citenamefont
  {Sarkar},\ and\ \citenamefont {Hemmer}}]{grotz_sensing_2011}%
  \BibitemOpen
  \bibfield  {author} {\bibinfo {author} {\bibfnamefont {B.}~\bibnamefont
  {Grotz}}, \bibinfo {author} {\bibfnamefont {J.}~\bibnamefont {Beck}},
  \bibinfo {author} {\bibfnamefont {P.}~\bibnamefont {Neumann}}, \bibinfo
  {author} {\bibfnamefont {B.}~\bibnamefont {Naydenov}}, \bibinfo {author}
  {\bibfnamefont {R.}~\bibnamefont {Reuter}}, \bibinfo {author} {\bibfnamefont
  {F.}~\bibnamefont {Reinhard}}, \bibinfo {author} {\bibfnamefont
  {F.}~\bibnamefont {Jelezko}}, \bibinfo {author} {\bibfnamefont
  {J.}~\bibnamefont {Wrachtrup}}, \bibinfo {author} {\bibfnamefont
  {D.}~\bibnamefont {Schweinfurth}}, \bibinfo {author} {\bibfnamefont
  {B.}~\bibnamefont {Sarkar}}, \ and\ \bibinfo {author} {\bibfnamefont
  {P.}~\bibnamefont {Hemmer}},\ }\href {\doibase 10.1088/1367-2630/13/5/055004}
  {\bibfield  {journal} {\bibinfo  {journal} {New Journal of Physics}\ }\textbf
  {\bibinfo {volume} {13}},\ \bibinfo {pages} {055004} (\bibinfo {year}
  {2011})}\BibitemShut {NoStop}%
\bibitem [{\citenamefont {Grinolds}\ \emph {et~al.}(2013)\citenamefont
  {Grinolds}, \citenamefont {Hong}, \citenamefont {Maletinsky}, \citenamefont
  {Luan}, \citenamefont {Lukin}, \citenamefont {Walsworth},\ and\ \citenamefont
  {Yacoby}}]{grinolds_nanoscale_2013}%
  \BibitemOpen
  \bibfield  {author} {\bibinfo {author} {\bibfnamefont {M.~S.}\ \bibnamefont
  {Grinolds}}, \bibinfo {author} {\bibfnamefont {S.}~\bibnamefont {Hong}},
  \bibinfo {author} {\bibfnamefont {P.}~\bibnamefont {Maletinsky}}, \bibinfo
  {author} {\bibfnamefont {L.}~\bibnamefont {Luan}}, \bibinfo {author}
  {\bibfnamefont {M.~D.}\ \bibnamefont {Lukin}}, \bibinfo {author}
  {\bibfnamefont {R.~L.}\ \bibnamefont {Walsworth}}, \ and\ \bibinfo {author}
  {\bibfnamefont {A.}~\bibnamefont {Yacoby}},\ }\href {\doibase
  10.1038/nphys2543} {\bibfield  {journal} {\bibinfo  {journal} {Nature
  Physics}\ }\textbf {\bibinfo {volume} {9}},\ \bibinfo {pages} {215} (\bibinfo
  {year} {2013})}\BibitemShut {NoStop}%
\bibitem [{\citenamefont {Gruber}\ \emph {et~al.}(1997)\citenamefont {Gruber},
  \citenamefont {Drabenstedt}, \citenamefont {Tietz}, \citenamefont {Fleury},
  \citenamefont {Wrachtrup},\ and\ \citenamefont
  {vonBorczyskowski}}]{gruber_scanning_1997}%
  \BibitemOpen
  \bibfield  {author} {\bibinfo {author} {\bibfnamefont {A.}~\bibnamefont
  {Gruber}}, \bibinfo {author} {\bibfnamefont {A.}~\bibnamefont {Drabenstedt}},
  \bibinfo {author} {\bibfnamefont {C.}~\bibnamefont {Tietz}}, \bibinfo
  {author} {\bibfnamefont {L.}~\bibnamefont {Fleury}}, \bibinfo {author}
  {\bibfnamefont {J.}~\bibnamefont {Wrachtrup}}, \ and\ \bibinfo {author}
  {\bibfnamefont {C.}~\bibnamefont {vonBorczyskowski}},\ }\href {\doibase
  10.1126/science.276.5321.2012} {\bibfield  {journal} {\bibinfo  {journal}
  {Science}\ }\textbf {\bibinfo {volume} {276}},\ \bibinfo {pages} {2012}
  (\bibinfo {year} {1997})}\BibitemShut {NoStop}%
\bibitem [{\citenamefont {Iizuka}\ \emph {et~al.}(2002)\citenamefont {Iizuka},
  \citenamefont {Watanabe}, \citenamefont {Sato},\ and\ \citenamefont
  {Nishikawa}}]{iizuka_millimeter-wave_2002}%
  \BibitemOpen
  \bibfield  {author} {\bibinfo {author} {\bibfnamefont {H.}~\bibnamefont
  {Iizuka}}, \bibinfo {author} {\bibfnamefont {T.}~\bibnamefont {Watanabe}},
  \bibinfo {author} {\bibfnamefont {K.}~\bibnamefont {Sato}}, \ and\ \bibinfo
  {author} {\bibfnamefont {K.}~\bibnamefont {Nishikawa}},\ }\href
  {http://search.ieice.org/bin/summary.php?id=e85-b_6_1169&category=B&year=2002&lang=E&abst=}
  {\bibfield  {journal} {\bibinfo  {journal} {IEICE TRANSACTIONS on
  Communications}\ }\textbf {\bibinfo {volume} {E85-B}},\ \bibinfo {pages}
  {1169} (\bibinfo {year} {2002})}\BibitemShut {NoStop}%
\bibitem [{\citenamefont {Mizuochi}\ \emph {et~al.}(2009)\citenamefont
  {Mizuochi}, \citenamefont {Neumann}, \citenamefont {Rempp}, \citenamefont
  {Beck}, \citenamefont {Jacques}, \citenamefont {Siyushev}, \citenamefont
  {Nakamura}, \citenamefont {Twitchen}, \citenamefont {Watanabe}, \citenamefont
  {Yamasaki}, \citenamefont {Jelezko},\ and\ \citenamefont
  {Wrachtrup}}]{mizuochi_coherence_2009}%
  \BibitemOpen
  \bibfield  {author} {\bibinfo {author} {\bibfnamefont {N.}~\bibnamefont
  {Mizuochi}}, \bibinfo {author} {\bibfnamefont {P.}~\bibnamefont {Neumann}},
  \bibinfo {author} {\bibfnamefont {F.}~\bibnamefont {Rempp}}, \bibinfo
  {author} {\bibfnamefont {J.}~\bibnamefont {Beck}}, \bibinfo {author}
  {\bibfnamefont {V.}~\bibnamefont {Jacques}}, \bibinfo {author} {\bibfnamefont
  {P.}~\bibnamefont {Siyushev}}, \bibinfo {author} {\bibfnamefont
  {K.}~\bibnamefont {Nakamura}}, \bibinfo {author} {\bibfnamefont
  {D.}~\bibnamefont {Twitchen}}, \bibinfo {author} {\bibfnamefont
  {H.}~\bibnamefont {Watanabe}}, \bibinfo {author} {\bibfnamefont
  {S.}~\bibnamefont {Yamasaki}}, \bibinfo {author} {\bibfnamefont
  {F.}~\bibnamefont {Jelezko}}, \ and\ \bibinfo {author} {\bibfnamefont
  {J.}~\bibnamefont {Wrachtrup}},\ }\href {\doibase 10.1103/PhysRevB.80.041201}
  {\bibfield  {journal} {\bibinfo  {journal} {Physical Review B}\ }\textbf
  {\bibinfo {volume} {80}},\ \bibinfo {pages} {041201} (\bibinfo {year}
  {2009})}\BibitemShut {NoStop}%
\bibitem [{\citenamefont {Pozar}(2004)}]{pozar2004microwave}%
  \BibitemOpen
  \bibfield  {author} {\bibinfo {author} {\bibfnamefont {D.}~\bibnamefont
  {Pozar}},\ }\href {http://books.google.co.uk/books?id=4wzpQwAACAAJ} {\emph
  {\bibinfo {title} {Microwave Engineering}}}\ (\bibinfo  {publisher} {Wiley},\
  \bibinfo {year} {2004})\BibitemShut {NoStop}%
\bibitem [{\citenamefont {Steiner}\ \emph {et~al.}(2010)\citenamefont
  {Steiner}, \citenamefont {Neumann}, \citenamefont {Beck}, \citenamefont
  {Jelezko},\ and\ \citenamefont {Wrachtrup}}]{steiner_universal_2010}%
  \BibitemOpen
  \bibfield  {author} {\bibinfo {author} {\bibfnamefont {M.}~\bibnamefont
  {Steiner}}, \bibinfo {author} {\bibfnamefont {P.}~\bibnamefont {Neumann}},
  \bibinfo {author} {\bibfnamefont {J.}~\bibnamefont {Beck}}, \bibinfo {author}
  {\bibfnamefont {F.}~\bibnamefont {Jelezko}}, \ and\ \bibinfo {author}
  {\bibfnamefont {J.}~\bibnamefont {Wrachtrup}},\ }\href {\doibase
  10.1103/PhysRevB.81.035205} {\bibfield  {journal} {\bibinfo  {journal}
  {Physical Review B}\ }\textbf {\bibinfo {volume} {81}},\ \bibinfo {pages}
  {035205} (\bibinfo {year} {2010})}\BibitemShut {NoStop}%
\bibitem [{\citenamefont {Neumann}\ \emph {et~al.}(2010)\citenamefont
  {Neumann}, \citenamefont {Beck}, \citenamefont {Steiner}, \citenamefont
  {Rempp}, \citenamefont {Fedder}, \citenamefont {Hemmer}, \citenamefont
  {Wrachtrup},\ and\ \citenamefont {Jelezko}}]{neumann_single-shot_2010}%
  \BibitemOpen
  \bibfield  {author} {\bibinfo {author} {\bibfnamefont {P.}~\bibnamefont
  {Neumann}}, \bibinfo {author} {\bibfnamefont {J.}~\bibnamefont {Beck}},
  \bibinfo {author} {\bibfnamefont {M.}~\bibnamefont {Steiner}}, \bibinfo
  {author} {\bibfnamefont {F.}~\bibnamefont {Rempp}}, \bibinfo {author}
  {\bibfnamefont {H.}~\bibnamefont {Fedder}}, \bibinfo {author} {\bibfnamefont
  {P.~R.}\ \bibnamefont {Hemmer}}, \bibinfo {author} {\bibfnamefont
  {J.}~\bibnamefont {Wrachtrup}}, \ and\ \bibinfo {author} {\bibfnamefont
  {F.}~\bibnamefont {Jelezko}},\ }\href {\doibase 10.1126/science.1189075}
  {\bibfield  {journal} {\bibinfo  {journal} {Science}\ }\textbf {\bibinfo
  {volume} {329}},\ \bibinfo {pages} {542} (\bibinfo {year}
  {2010})}\BibitemShut {NoStop}%
\bibitem [{\citenamefont {Stepanov}\ \emph {et~al.}(2015)\citenamefont
  {Stepanov}, \citenamefont {Cho}, \citenamefont {Abeywardana},\ and\
  \citenamefont {Takahashi}}]{stepanov_high-frequency_2015}%
  \BibitemOpen
  \bibfield  {author} {\bibinfo {author} {\bibfnamefont {V.}~\bibnamefont
  {Stepanov}}, \bibinfo {author} {\bibfnamefont {F.~H.}\ \bibnamefont {Cho}},
  \bibinfo {author} {\bibfnamefont {C.}~\bibnamefont {Abeywardana}}, \ and\
  \bibinfo {author} {\bibfnamefont {S.}~\bibnamefont {Takahashi}},\ }\href
  {\doibase 10.1063/1.4908528} {\bibfield  {journal} {\bibinfo  {journal}
  {Applied Physics Letters}\ }\textbf {\bibinfo {volume} {106}},\ \bibinfo
  {pages} {063111} (\bibinfo {year} {2015})}\BibitemShut {NoStop}%
\end{thebibliography}%

\end{document}